\documentclass[preprint,12pt]{aastex}

\shorttitle{\emph{SMA} Observations of NGC 7538}
\shortauthors{Qiu et al.}
\begin{document}

\title{Outflows, Accretion, and Clustered Protostellar Cores
around a Forming O Star}

\author{Keping Qiu}
\affil{Department of Astronomy, Nanjing University, Nanjing 210093,
China} \affil{Harvard-Smithsonian Center for Astrophysics, 60 Garden
Street, Cambridge, MA, USA} \affil{Max-Planck-Institute for
Radioastronomy, Auf dem H\"{u}gel 69, 53121 Bonn, Germany}
\email{kqiu@mpifr-bonn.mpg.de}

\author{Qizhou Zhang}
\affil{Harvard-Smithsonian Center for Astrophysics, 60 Garden
Street, Cambridge, MA, USA} \email{qzhang@cfa.harvard.edu}

\author{Karl M. Menten}
\affil{Max-Planck-Institute for Radioastronomy, Auf dem H\"{u}gel
69, 53121 Bonn, Germany} \email{kmenten@mpifr-bonn.mpg.de}

\begin{abstract}
We present a Submillimeter Array study in the 1.3 mm waveband of the
NGC 7538 IRS 1--3 massive star-forming region. The brightest core in
the mm continuum map, MM1, harbors the IRS 1 young O star. The core
has a gas temperature of about 245 K and shows spatially unresolved
emission in complex organic molecules, all typical of a hot
molecular core. Toward MM1, redshifted absorption is seen in
molecular lines with different energies above the ground state. This
absorption probes inward motion of the dense gas toward the central
young O star, and the estimated mass accretion rate reaches
$10^{-3}$ $M_{\odot}$\,yr$^{-1}$. Multiple outflows are seen in the
CO and $^{13}$CO maps. The gas mass of 50 $M_{\odot}$ and mass
outflow rate of $2.5\times10^{-3}$ $M_{\odot}$\,yr$^{-1}$ measured
in CO line wings are dominated by the MM1 outflow, which is most
likely driven by a fast wide-angle wind. Apart from MM1, we discover
eight new dusty cores, MM2--9, within a projected distance of 0.35
pc from MM1. These cores show no counterpart in infrared or radio
continuum emission, while seven of them appear to be forming
intermediate- to high-mass stars. This manifests a deeply embedded
star-forming component of the parent cloud of IRS 1--3. Apparently
we are observing a Trapezium system in formation, and the system is
presumably surrounded by a cluster of lower mass stars.

\end{abstract}


\section{Introduction} \label{intro}
NGC 7538 is a large, optically visible H{\scriptsize II} region in
the Perseus arm. About 2 arcmin (1.5 pc) southeast to the center of
the H{\scriptsize II} region is a site of active massive star
formation known for three radio and infrared (IR) sources IRS 1--3
\citep{Martin73, Wynn74}. Located at a distance of 2.65 kpc
\citep{Moscadelli09}, the IRS 1--3 region has a total luminosity of
$1.4\times10^5$ $L_{\odot}$, of which the luminosity of IRS 1 is
about $8\times10^4$ $L_{\odot}$, equivalent to a ZAMS spectral type
of O7.5 \citep{Werner79, Campbell84b, Akabane05}. IRS 2 and IRS 3
are two nearby sources less obscured than IRS 1. IRS 2 has a
luminosity of $5\times10^4$ $L_{\odot}$ and is associated with a
compact H{\scriptsize II} region; IRS 3 has a luminosity of
$6\times10^3$ $L_{\odot}$ and is associated with an ultracompact
H{\scriptsize II} region \citep{Campbell84b, Bloomer98, Sandell09}.

IRS 1 has attracted numerous studies. Radio continuum observations
show that the ionized gas has a double-lobed structure within the
central $0.\!''2$ region and more extended halos at a $1''$ scale
\citep{Campbell84a, Gaume95, Sandell09}, and exhibits time variation
in emission flux \citep{Franco04}. Radio recombination lines (RRLs)
observed at cm and mm wavelengths show extremely broad line widths,
suggestive of expanding motion of the ionized gas \citep{Gaume95,
Keto08}. In high-angular-resolution mid-IR images, a shell-like
structure extending from IRS 1 to about $4''$ to the northwest is
interpreted as dust emission on the wall of the outflow cavity
\citep{DeBuizer05}. $K'$-band images obtained with speckle
interferometry show a fan-shaped nebula extending from IRS 1 toward
the northwest, and the emission is predominantly attributed to
scattered light from an outflow cavity \citep{Kraus06}. On larger
scales, single-dish CO observations reveal a bipolar outflow with a
northwest-southeast (NW-SE) orientation which is centered at IRS 1
\citep{Kameya89, Davis98}. IRS 1 is associated with various types of
molecular masers \citep[e.g.,][and references therein]{Pratap89}.
Pestalozzi et al. (2004) modeled a group of methanol masers detected
toward IRS 1 as tracing an edge-on Keplerian disk around a central
star of 30 $M_{\odot}$, although the orientation of a possible disk
around IRS 1 is controversial \citep[e.g.,][]{DeBuizer05,
Klaassen09}.

Existing high-angular-resolution observations toward NGC 7538 IRS 1
mostly focus on hot dust and ionized gas in the close vicinity of
the young O star, while kinematics and physical conditions of
molecular gas in the dusty envelope is less clear. Moreover, the
molecular outflow in this region has not been observed at high
angular resolutions; a pioneering interferometric study performed by
Scoville et al. (1986) is of moderate resolution and image quality.
In this paper we present a 1.3 mm waveband study of the NGC 7538 IRS
1 region using the Submillimeter Array\footnote{The Submillimeter
Array is a joint project between the Smithsonian Astrophysical
Observatory and the Academia Sinica Institute of Astronomy and
Astrophysics and is funded by the Smithsonian Institution and the
Academia Sinica.}, aimed at understanding physical conditions and
kinematics of the gas and dust cocoon surrounding the young O star
and the nature of the molecular outflow. In its course we discovered
clustered star-forming cores deeply embedded within the parent cloud
of IRS 1--3 and multiple molecular outflows. A description of
observations and data reduction is presented in Section \ref{obs}.
Observational results are presented in Section \ref{result},
followed by discussions on their implications in Section \ref{dis}.
Finally, a summary of the main findings is given in Section
\ref{conclu}.

\section{Observations and Data Reduction} \label{obs}
\subsection{SMA Observations}
The SMA observations were undertaken on 2007 July 13 with eight
antennas in the Compact-north configuration. The weather was very
good during the observation, with 1.5 to 2.5 mm precipitable water
vapor (PWV), corresponding to atmospheric optical depth at 225 GHz,
${\tau}_{225\mathrm {GHz}}$, around 0.08 to 0.12. To cover the
entire outflow we observed two fields centered at (R.A.,
decl.)$_{\mathrm J2000}=(23^{\mathrm h}13^{\mathrm m}43.\!^{\mathrm
s}75, 61^{\circ}28{'}21.\!{''}5)$ and $(23^{\mathrm h}13^{\mathrm
m}46.\!^{\mathrm s}75, 61^{\circ}27{'}59.\!{''}5)$, guided by
previous single-dish observations \citep{Davis98}. The 2$\times$2
GHz correlator was configured to cover rest frequencies about 219.2
to 221.2 GHz in the lower sideband and about 229.2 to 231.2 GHz in
the upper sideband, with a uniform spectral resolution of 812.5 kHz
($\sim$1.1 km\,s$^{-1}$). We observed 3c273 and Callisto for
bandpass calibration and MWC349, for which a flux of 1.65 Jy at 225
GHz was adopted, for flux calibration. Time dependent gains were
monitored by interleavingly observing J0102+584 and BL Lac every 30
mins. The primary calibrations were carried out with the IDL MIR
package, which allows to determine antenna gains (versus frequency
or time) by simultaneously fitting multiple calibrators. The
calibrated visibilities were exported to MIRIAD for further
processing. A continuum database was constructed from line-free
channels using the MIRIAD task UVLIN. We self-calibrated the
continuum solving for phase corrections with decreasing time
intervals (5, 2, 1 min), and then solving for phase and amplitude
corrections for the last iteration. The solutions from the
self-calibration were applied to spectral line data as well.

\begin{figure}
\epsscale{1} \plotone{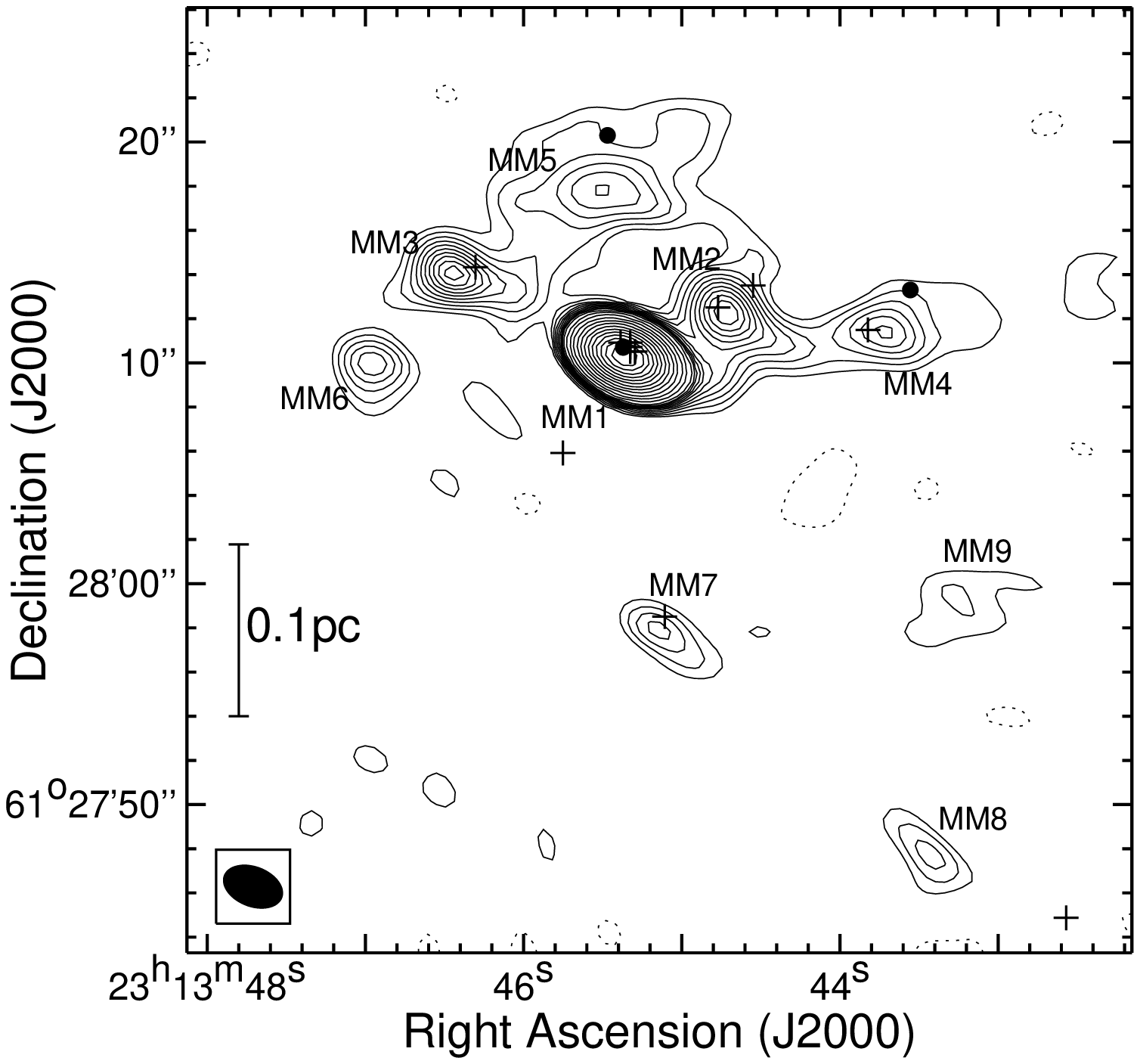} \caption{1.3 mm continuum emission
shown in solid contours with contour levels increasing from 30 to
165 mJy\,beam$^{-1}$ in steps of 15 mJy\,beam$^{-1}$ and continuing
to 3255 mJy\,beam$^{-1}$ in steps of $30\times(1, 2, 3, ..., 12)$
mJy\,beam$^{-1}$; dotted contours show negative emission with
absolute levels the same as that of the positive. The plus symbols
depict water maser spots from Kameya et al. (1990); three filled
dots mark the positions of IRS 1--3, where IRS 1 roughly coincides
with MM1, IRS 2 is about $10''$ to the north and IRS 3 about $14''$
to the west. Hereafter a filled ellipse in the lower left delineates
the synthesized beam at FWHM. \label{cont_map}}
\end{figure}

In Figure \ref{cont_map}, the predominant core MM1 is more than 10
times brighter than MM2--9. As the dynamical range of an
interferometric image is limited by incomplete $(u,v)$ sampling, we
assess the reliability of MM2--9. We imaged J0102+584 and BL Lac and
found that each of the two quasars appears as a point source at the
phase center, and the position agrees with the nominal coordinate to
$<0.1''$. We also found that features associated with MM2--9 are
identifiable in the continuum and spectral line images without
self-calibration. Furthermore, we obtained a model of MM1 by
restricting CLEAN (a deconvolution task in MIRIAD) to a polygon
closely around MM1, subtracted ``MM1'' from the calibrated
visibilities using the task UVMODEL, and imaged the residual
visibilities; MM1 was completely invisible while MM2--9 were clearly
seen in the image, indicating that none of the fainter cores was
falsely generated due to sidelobes of MM1 during deconvolution.

The final maps have a synthesized beam of $\sim$$3''\times2''$, and
r.m.s. sensitivities of $\sim$5 mJy for continuum and $\sim$35 mJy
per 1.2 km\,s$^{-1}$ for lines.

\subsection{Short Spacings in CO and $^{13}$CO (2--1)}
In an interferometric observation, structures more extended than
1.2$\lambda/b_{\mathrm {min}}$, where $\lambda$ is the observing
wavelength and $b_{\mathrm {min}}$ the shortest length of the
projected baselines, cannot be effectively sampled. This filtering
effect is significant for CO and $^{13}$CO maps, particularly at
velocities around the cloud velocity. Adding in single-dish
observations is a way to overcome this effect
\citep[e.g.,][]{Qiu09b}. For CO (2--1), we combined the SMA data
with data obtained from the James Clerk Maxwell Telescope (JCMT) by
Davis et al. (1998). For $^{13}$CO (2--1), we carried out on-the-fly
mapping observations with the Caltech Submillimeter
Observatory\footnote[2]{The Caltech Submillimeter Observatory is
supported by the NSF grant AST-0229008} (CSO) 10.4 m telescope on
2008 June 11. During the last observation, the weather was
reasonably good with about 4 mm PWV and ${\tau}_{225\mathrm {GHz}}$
around 0.2. We obtained a $13\times13$ grid map centered on (R.A.,
decl.)$_{\mathrm J2000}=(23^{\mathrm h}13^{\mathrm m}43.\!^{\mathrm
s}8, 61^{\circ}28{'}22.\!{''}0)$. The grid cell spacing is 10$''$,
roughly $1/3$ of the FWHM beam at this frequency. The effective
integration time toward each cell is about 10 s. The AOS
spectrometer has 1024 channels throughout the 50 MHz bandwidth, thus
a spectral resolution of 48.8 kHz (or 0.066 km\,s$^{-1}$). The data
were reduced using the standard CLASS package. The combination of
the SMA and single-dish data was conducted in MIRIAD following a
procedure outlined in Zhang et al. (1995). The final CO and
$^{13}$CO maps were smoothed to 2 km\,s$^{-1}$ per channel.

\section{Results} \label{result}

\subsection{1.3 mm Continuum Emission} \label{cont}
Figure \ref{cont_map} presents the 1.3 mm continuum map, which
reveals a cluster of nine mm cores, namely MM1 to MM9 in order of
peak fluxes. Measured parameters of these cores are listed in Table
\ref{mm_tbl}. The brightest core MM1 coincides with the luminous IR
and radio source IRS 1. The 1.3 mm continuum flux has a contribution
of $\sim$1.5 Jy from the free-free emission extrapolated from cm
measurements \citep{Sandell09}, and the dust emission flux is about
2.1 Jy. The 1.3 mm emission from MM2 to MM9 is entirely dominated by
the dust emission, since none of them is seen in cm continuum. In
contrast to IRS 1, neither IRS 2 nor IRS 3 appears to be embedded
within a dense dusty core.

Assuming thermal equilibrium between the dense gas and dust, we
adopt a temperature of 245 K for MM1 derived from a local
thermodynamical equilibrium (LTE) model of the CH$_3$CN emission
(Section \ref{mm1_line}). Cores MM2--9 have much lower temperatures;
we estimate temperatures from NH$_3$ data \citep{Zheng01} for MM6--9
(Section \ref{newmm_line}) and adopt an averaged temperature of 40 K
for MM2--5 (Section \ref{cluster}). We then estimate gas masses of
all the cores by adopting the dust opacity $\kappa_{\mathrm
{225GHz}}=1$ cm$^2$\,g$^{-1}$ \citep{Ossenkopf94}, which is
equivalent to an opacity index $\beta$ of 1.375 for $\kappa_{\mathrm
{\nu}}=10\,(\frac{\nu}{\mathrm {1.2 THz}}$)$^{\beta}$
cm$^2$\,g$^{-1}$ \citep{Hildebrand83}, and a canonical gas-to-dust
mass ratio of 100. The derived core masses are listed in Table
\ref{mm_tbl}. The error bars of the core masses presented in
Table \ref{mm_tbl} account for errors in measuring the continuum
fluxes and calculating the temperatures. For MM1, the uncertainty in
extrapolating the free-free emission is also taken into account. For
MM2--5, an averaged uncertainty of $35\%$ is assumed for the adopted
temperature. There is a systemic uncertainty, $9\%$, introduced by
the distance error \citep[0.12 kpc,][]{Moscadelli09}. Other
uncertainties come from possible variations of the gas-to-dust mass
ratio and the dust opacity; the latter can be constrained with
future high angular resolution observations at different
wavelengths.

\subsection{Molecular Outflows} \label{outflow}
Figure \ref{co_int} shows the velocity integrated CO (2--1) emission
obtained from the combined SMA and JCMT data sets. Velocity channel
maps of the emission are shown in Figure \ref{co_chan}. Overall the
CO emission shows spatially separated blue- and redshfited lobes in
a NW-SE orientation, whereas the detailed structures are ascribed to
multiple outflows rather than a simple bipolar outflow seen in
previous single-dish observations.

In the integrated map, the blueshifted lobe appears as a curved
filament originating from MM1 and diverging into three ``branches''
toward the northwest, north, and northeast. However, from the
orientation, the northeastern branch with a position angle (PA) of
about 40$^{\circ}$ is more likely arising from MM4 or IRS 3; from
the channel maps at $-78$ to $-64$ km\,s$^{-1}$, the northern branch
with a PA of about 0$^{\circ}$ may originate from IRS 3 or MM4 as
well. In addition, at least four more dusty cores, MM2, MM6, MM7,
and MM8 appear to be associated with blueshifted CO outflows: an
elongated feature extending from MM6 to the southeast is seen in the
integrated map and blueshifted channel maps of $\geq$$-78$
km\,s$^{-1}$; emission around MM8 and extending to the northwest is
seen in the integrated map and blueshifted channel maps of
$\geq$$-74$ km\,s$^{-1}$; emission associated with MM2 is seen as a
knot to its immediate southwest at $-78$ to $-72$ km\,s$^{-1}$ and
as a compact elongation at $-70$ to $-68$ km\,s$^{-1}$; compact and
faint emission extending from MM7 to the southwest is visible at
$-70$ to $-66$ km\,s$^{-1}$.

The redshifted emission appears as a large-scale ``arc'' in the
integrated map, while the detailed structures can be attributed to
at least two cores, MM1 and MM7. MM1 is presumably the powering
source of the emission extending from around MM1 to a bright clump
about $20''$ (0.26 pc) to the southeast; the clump is encompassed by
a C-shaped structure seen in C$^{18}$O (2--1). MM7 is likely the
central source of the CO emission in its close vicinity: a knot
about $7''$ (0.09 pc) to the northeast seen at $-38$ and $-36$
km\,s$^{-1}$ is most likely the redshifted counterpart of the
compact feature seen at $-70$ to $-66$ km\,s$^{-1}$; a southwestern
clump identifiable in the integrated map and channel maps of
$\geq$$-48$ km\,s$^{-1}$ is probably associated with MM7 as well.

Figures \ref{13co_int} and \ref{13co_chan} show the velocity
integrated and channel maps, respectively, of the $^{13}$CO (2--1)
emission, obtained from the combined SMA and CSO data. In general
the $^{13}$CO emission shows a morphology similar to that of the CO
emission, but within a narrower velocity range. The MM1 NW-SE
outflow appears more predominant in the $^{13}$CO emission. Although
fainter, the PA$\sim40^{\circ}$ and $0^{\circ}$ branches of the
blueshifed emission are both identifiable. The redshifted emission
of the MM7 outflow is clearly seen; in particular in the channel map
of $-48$ km\,s$^{-1}$ it shows a bi-conical structure.

A schematic view of the multiple outflows detected here is shown in
Figure \ref{sketchy}, which presents a tentative anatomy of the
complicated outflow structures seen in CO and $^{13}$CO (2--1). A
more detailed discussion on the MM1 outflow is presented in Section
\ref{dis_outflow}.

Following Qiu et al. (2009), the outflow parameters, including the
mass ($M_{\mathrm {out}}$), momentum ($P_{\mathrm {out}}$), energy
($E_{\mathrm {out}}$), dynamical timescale ($t_{\mathrm {dyn}}$),
mass outflow rate ($\dot{M}_{\mathrm {out}}$) and momentum outflow
rate ($\dot{P}_{\mathrm {out}}$), are calculated. The optical depth
effect as a function of velocity and spatial distribution is
corrected by comparing the CO with $^{13}$CO fluxes and adopting a
C-to-$^{13}$C ratio of 79 \citep{Wilson94}. The derived parameters
are listed in Table \ref{outflow_tbl}. To avoid contamination from
ambient diffuse gas, we conservatively calculate the mass of the
outflowing gas at velocities of $\leq$$-64$ and $\geq$$-48$
km\,s$^{-1}$, which are more than 6 km\,s$^{-1}$ away from the
systemic velocity of $-57.4$ km\,s$^{-1}$ \citep{vanderTak00}.
However, from Figure \ref{co_chan}, the emission at $-62$ and $-50$
km\,s$^{-1}$ has significant contribution from outflows. If we
choose the lowest outflow velocities to be $-62$ and $-50$
km\,s$^{-1}$, the gas mass amounts to 110 $M_{\odot}$. Hence, the
total outflow mass should be in the range of 50 to 100 $M_{\odot}$.

\subsection{High-density Tracing Molecular Lines}

\subsubsection{Emission and absorption toward MM1} \label{mm1_line}
MM1 shows molecular chemistry typical of a hot molecular core
\citep[HMC,][]{Cesaroni05,Beuther09}. Excluding CO and $^{13}$CO
(2--1), around 70 lines arising from 15 molecular species are
detected toward MM1 with peak fluxes above 6$\sigma$ (Table
\ref{line_tbl}). Spatially unresolved emission from complex
O-bearing molecules, e.g., methyl formate (CH$_3$OCHO) and ethanol
(C$_2$H$_5$OH), and N-bearing molecules, e.g., methyl cyanide
(CH$_3$CN) and isocyanic acid (HNCO), is clearly seen. In
particular, the number of detected lines is largely dominated by
O-bearing molecules, e.g., CH$_3$OH, CH$_3$OCHO, and C$_2$H$_5$OH.

The CH$_3$CN emission is an excellent temperature and density probe
\citep{Araya05,Cesaroni05}. The  CH$_3$CN lines detected in MM1
cover a wide range of upper level energies above the ground state,
$E_{\mathrm {up}}$ (69--526 K). Following Qiu \& Zhang (2009), we
perform a simultaneous fitting to all the detected $K$ components of
the CH$_3$CN (12--11) emission, where the temperature, column
density, and source size are determined by means of grid search
$\chi$-square minimization. In Figure \ref{ch3cn}, the best fit
model agrees very well with the observation, and yields a
temperature of 245$^{+25}_{-25}$ K, a column density of
5$^{+1}_{-1}\times$10$^{16}$ cm$^{-2}$, and a source size of
1450$^{+170}_{-130}$ AU. As the fitting is highly nonlinear,
uncertainties are determined from a Monte Carlo analysis accounting
for the 1$\sigma$ noise and the $\sim$10\% uncertainty of flux
calibration. Compared to the rotation diagram method, the fitting
improves the parameter derivation by solving for the optical depths
\citep{Goldsmith99}, which range from 0.16 for the $K$=8 line to
2.99 for the $K$=3 line in the best fit model. The fitting, however,
simply assumes uniform gas. Density and temperature gradients
certainly exist in a gas and dust envelope surrounding a (proto)star
\citep[e.g.,][]{Scoville76, Osorio09}. The best fit model thus
represents a characteristic estimate. The beam-averaged CH$_3$CN
column density, 3$\times$10$^{15}$ cm$^{-2}$, can be compared to the
H$_2$ column density (1.4$\times$10$^{24}$ cm$^{-2}$, derived from
the dust continuum peak) for an estimate of the CH$_3$CN abundance,
which is  2.1$\times$$10^{-9}$, in agreement with typical values of
$10^{-10}$--$10^{-8}$ measured in other HMCs
\citep[e.g.,][]{Hatchell98, Chen06, Zhang07}. Finally, the
C$_2$H$_5$OH (13$_{1,13}$--12$_{0,12}$) (${\nu}_0$=220601.93 MHz)
line in Figure 7 is often identified as CH$_3$$^{13}$CN
(12$_3$--11$_3$) (${\nu}_0$=220599.98 MHz) in the literature, which
would imply a CH$_3$CN (12$_3$--11$_3$) optical depth of $\sim$42,
abnormally high for such a high-density and warm gas tracer. If the
optical depths were that high the low $K$ components of the CH$_3$CN
(12--11) emission would be entirely optically thick and show
flattened line peaks; this clearly disagrees with the observed
spectra. Here we believe the emission arising from C$_2$H$_5$OH
(13$_{1,13}$--12$_{0,12}$).

A secondary emission feature around $-$50 km\,s$^{-1}$ is seen in
Nitrogen- and Sulfur-bearing species, i.e., CH$_3$CN, HNCO, OCS, SO,
and SO$_2$. The feature is not detected ($<6\sigma$) in lines of
H$_2^{13}$CO, CH$_3$OH, or any other large O-bearing molecules. The
emission is probably arising from an interaction between the
outflowing material and a clump residing in the far side of the
central young O star. The lines in Nitrogen- and Sulfur-bearing
species could be related to shock-induced chemistry, while the
deficiency in lines of CH$_3$OH, H$_2^{13}$CO, and larger O-bearing
molecules could be due to an initially low abundance of the ``key''
species CH$_3$OH and H$_2$CO for the clump
\citep[e.g.,][]{Charnley92}. The existing observations cannot shed
further light on this issue.

Thought to be a probe of inward motion of dense gas surrounding a
forming star, molecular lines with redshifted absorption observed at
high-angular-resolutions have been reported in a very few high-mass
star-forming cores \citep[e.g.,][]{Zhang97, Beltran06}. Toward MM1,
an inverse P-Cygni profile with redshifted absorption
($<$$-6\sigma$) is seen in lines of C$^{18}$O (2--1), SO
(5$_6$--4$_5$), HNCO (10$_{0,10}$--9$_{0,9}$), CH$_3$OH
(8$_{-1,8}$--7$_0,7$)$E$ and (8$_{0,8}$--7$_1,6$)$E$, and CH$_3$CN
(12--11) $K$=2,3 (see, e.g., Figure \ref{absorption}). In these
lines, the absorption comes from dense gas lying on the near side of
the central continuum source, absorbing bright continuum emission
and moving toward the source (away from the observer). Meanwhile,
the emission is attributed to dense gas residing on the far side of
the central source and moving toward the observer; within a
$3''\times2''$ beam, dense gas with velocities largely along the
plane of sky (i.e., orthogonal to the line of sight) can contribute
to the emission as well. The lines showing redshifted absorption
have $E_{\mathrm {up}}$ spanning a range of 16--133 K. Roughly
speaking, the absorption in lines of higher excitation (e.g., the
CH$_3$OH and CH$_3$CN lines) may trace inner, warmer gas with
velocities mostly along the line of sight, hence appears more biased
toward redshifted velocities. In addition, analytical models of a
dense core under self-similar gravitational collapse shows that the
infall velocity scales with the radius as $r^{-0.5}$
\citep{Larson72, Shu77}, which may partly contribute to the redward
shift of the absorption in lines of higher $E_{\mathrm {up}}$.
Assuming the collapse proceeds in a spherical geometry, the
accretion rate can be estimated following
$\dot{M}_{acc}=4{\pi}r^2{\rho}V_{infall}$, where $\dot{M}_{acc}$ is
the mass accretion rate, $r$ the radius of the infalling envelope,
$\rho$ the mass density, and $V_{infall}$ the infall velocity. From
the dust continuum peak the H$_2$ number density is about $10^7$
cm$^{-3}$ averaged over $r\sim3000$ AU. The measured infall velocity
is about 2.5 km\,s$^{-1}$. As an order-of-magnitude estimate,
$\dot{M}_{acc}$ amounts to $3\times10^{-3}$ $M_{\odot}\,{\rm
yr}^{-1}$. However, the accretion may considerably deviate from a
spherical geometry, in particular for the inner part closely
surrounding the central source. The free-free emission in IRS 1
shows a double-lobed morphology at the center, suggesting that the
accretion proceeds along a flattened structure perpendicular to the
ionized gas elongation. If the collapse occurs inside a solid angle
${\it \Omega}$, $\dot{M}_{acc}$ scales to ${\it
\Omega}/(4\pi)\cdot3\times10^{-3}$ $M_{\odot}\,{\rm yr}^{-1}$.

\subsubsection{Emission from new cores MM2 to MM9} \label{newmm_line}
In addition to MM1, the dust continuum map reveals eight new cores,
MM2--9. The C$^{18}$O (2--1), SO (5$_6$--4$_5$), and CH$_3$OH
(8$_{-1,8}$--7$_{0,7}$) {\it E} lines show emission from part of
these new cores and extended structures, and are not significantly
affected by outflows or ambient diffuse gas. Figure \ref{mom0} shows
moment-0 (velocity averaged) maps of these lines.

\begin{figure}
\epsscale{1} \plotone{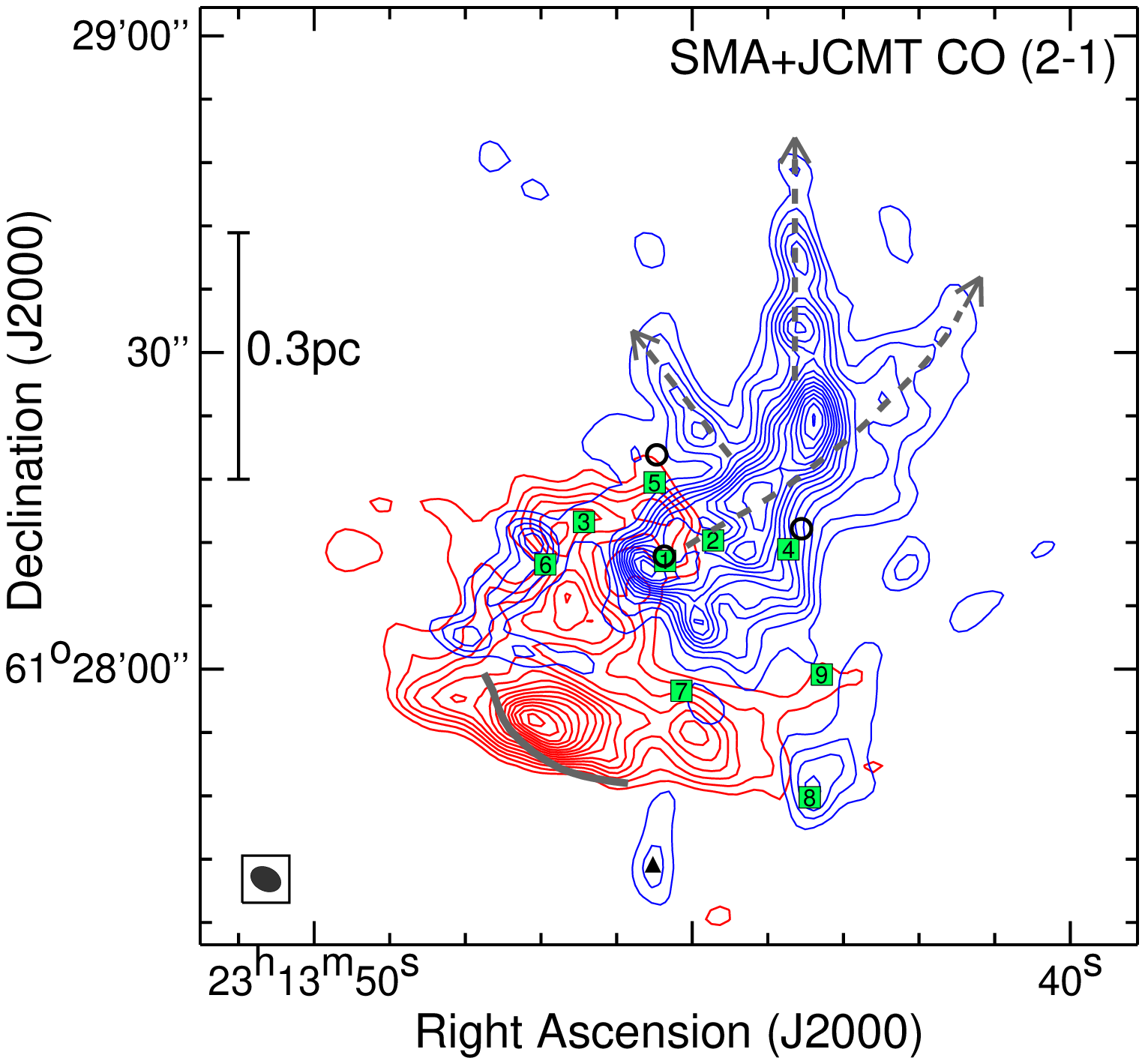} \caption{Integrated CO (2--1) emission
obtained from the combined SMA and JCMT data. Blue contours show
emission integrated from $-78$ to $-64$ km\,s$^{-1}$ and the contour
levels start from 10.2 Jy$\cdot$km\,s$^{-1}$ and increase in steps
of 6.8 Jy$\cdot$km\,s$^{-1}$; red contours show emission integrated
from $-48$ to $-30$ km\,s$^{-1}$ and the contour levels start from
20.2 Jy$\cdot$km\,s$^{-1}$ and increase in steps of 10.1
Jy$\cdot$km\,s$^{-1}$. The numbered squares mark dusty cores MM1--9
(see Figure \ref{cont_map}); three thick circles denote IRS 1--3; a
filled triangle marks a newly discovered 229.8 GHz CH$_3$OH maser
(see Section \ref{newmm_line} and Figures \ref{mom0}c, \ref{maser}).
A gray curve delineates a C-shaped structure seen in C$^{18}$O
(2--1) (see Section \ref{newmm_line} and Figure \ref{mom0}a), and
dashed arrows outline the three ``branches'' of the blueshifted
outflow (see Section \ref{outflow}). \label{co_int}}
\end{figure}

\begin{figure}
\epsscale{1} \plotone{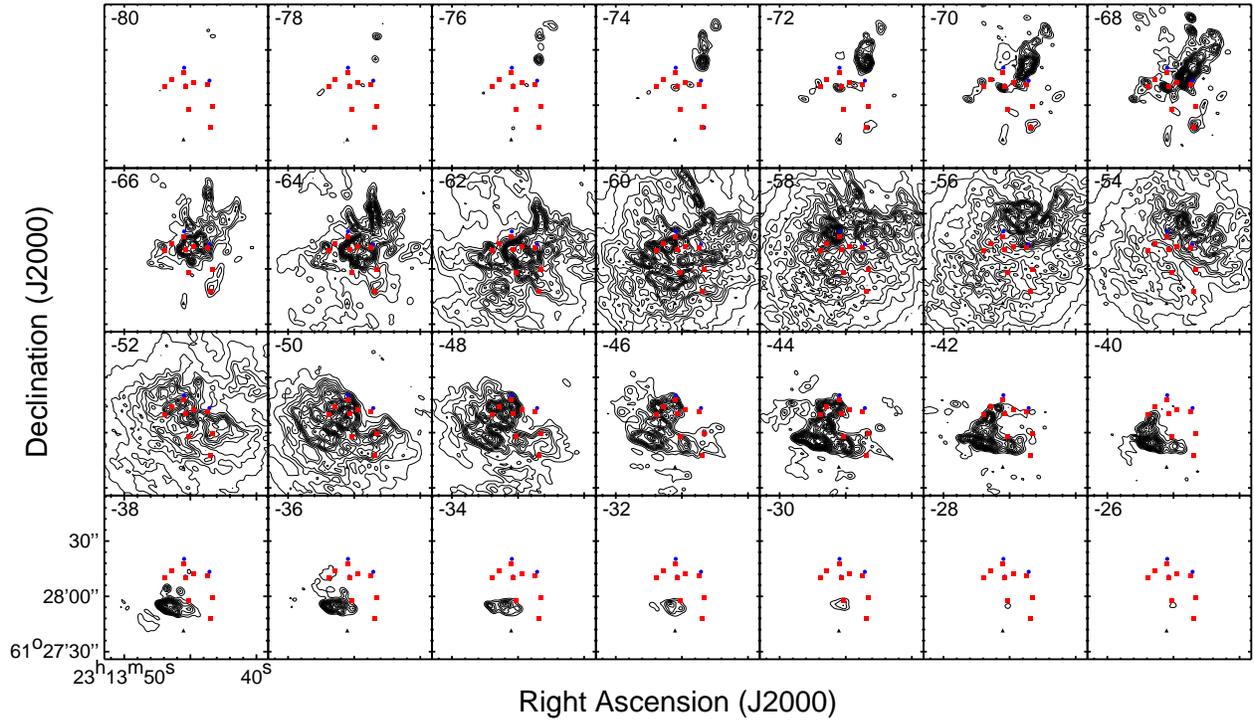} \caption{Velocity channel maps of CO
(2--1) emission obtained from the combined SMA and JCMT data. The
starting and spacing contour levels are 1.2 Jy\,beam$^{-1}$ for
channels from $-66$ to $-46$ km\,s$^{-1}$ and 0.6 Jy\,beam$^{-1}$ in
outer line wings. The central velocity of each channel is shown in
the upper left of each panel; cores MM1--9 and IR sources IRS 1--3
are denoted as filled squares and dots, respectively.
\label{co_chan}}
\end{figure}

\begin{figure}
\epsscale{1} \plotone{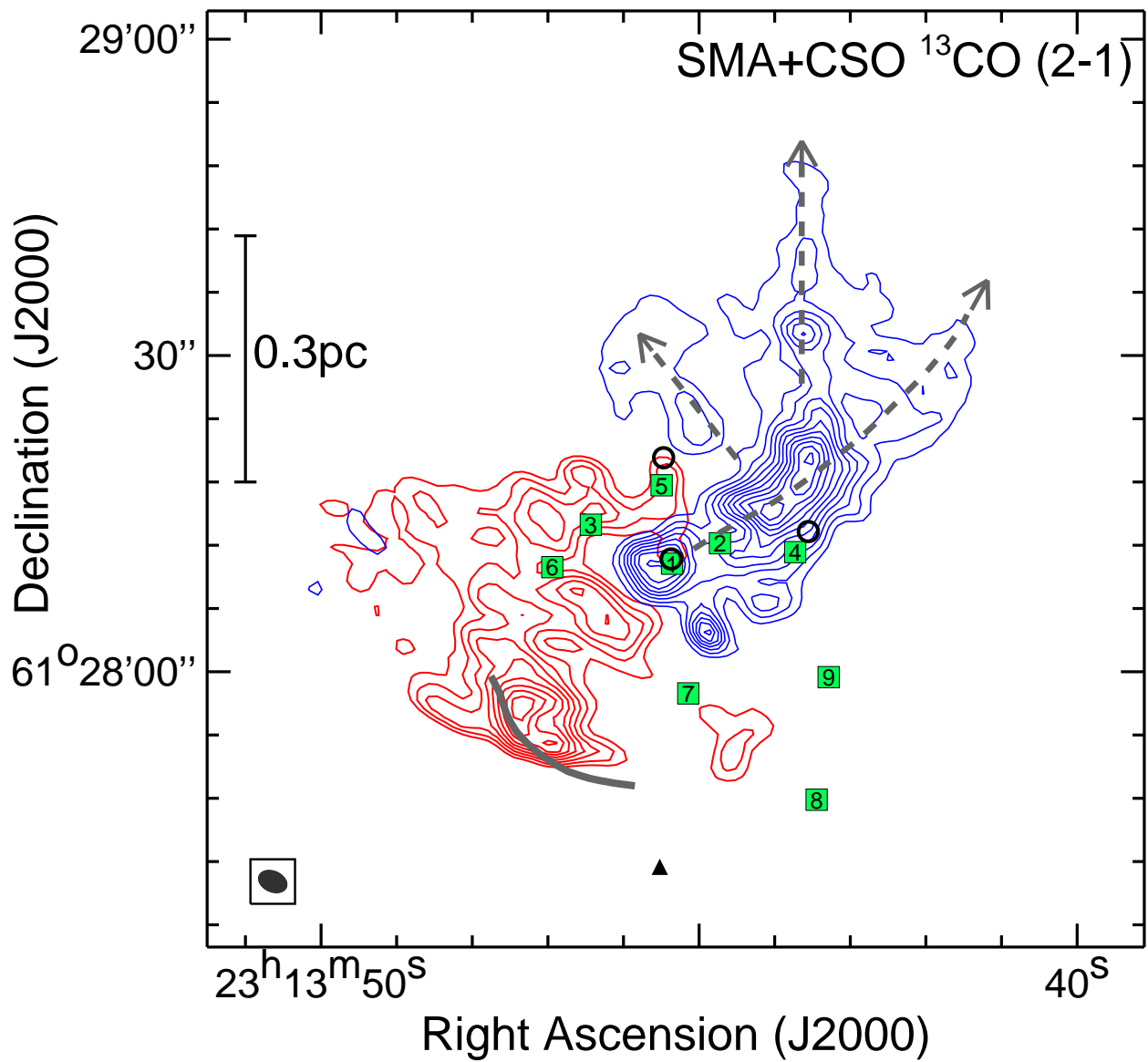} \caption{Integrated $^{13}$CO (2--1)
emission obtained from the combined SMA and CSO data. Blue contours
show emission integrated from $-72$ to $-64$ km\,s$^{-1}$ and the
contour levels start from 2.6 Jy$\cdot$km\,s$^{-1}$ and increase in
steps of 1.3 Jy$\cdot$km\,s$^{-1}$; red contours show emission
integrated from $-48$ to $-42$ km\,s$^{-1}$ and the contour levels
start from 3.6 Jy$\cdot$km\,s$^{-1}$ and increase in steps of 1.2
Jy$\cdot$km\,s$^{-1}$. Symbols are the same as those in Figure
\ref{co_int}. \label{13co_int}}
\end{figure}

\begin{figure}
\epsscale{1} \plotone{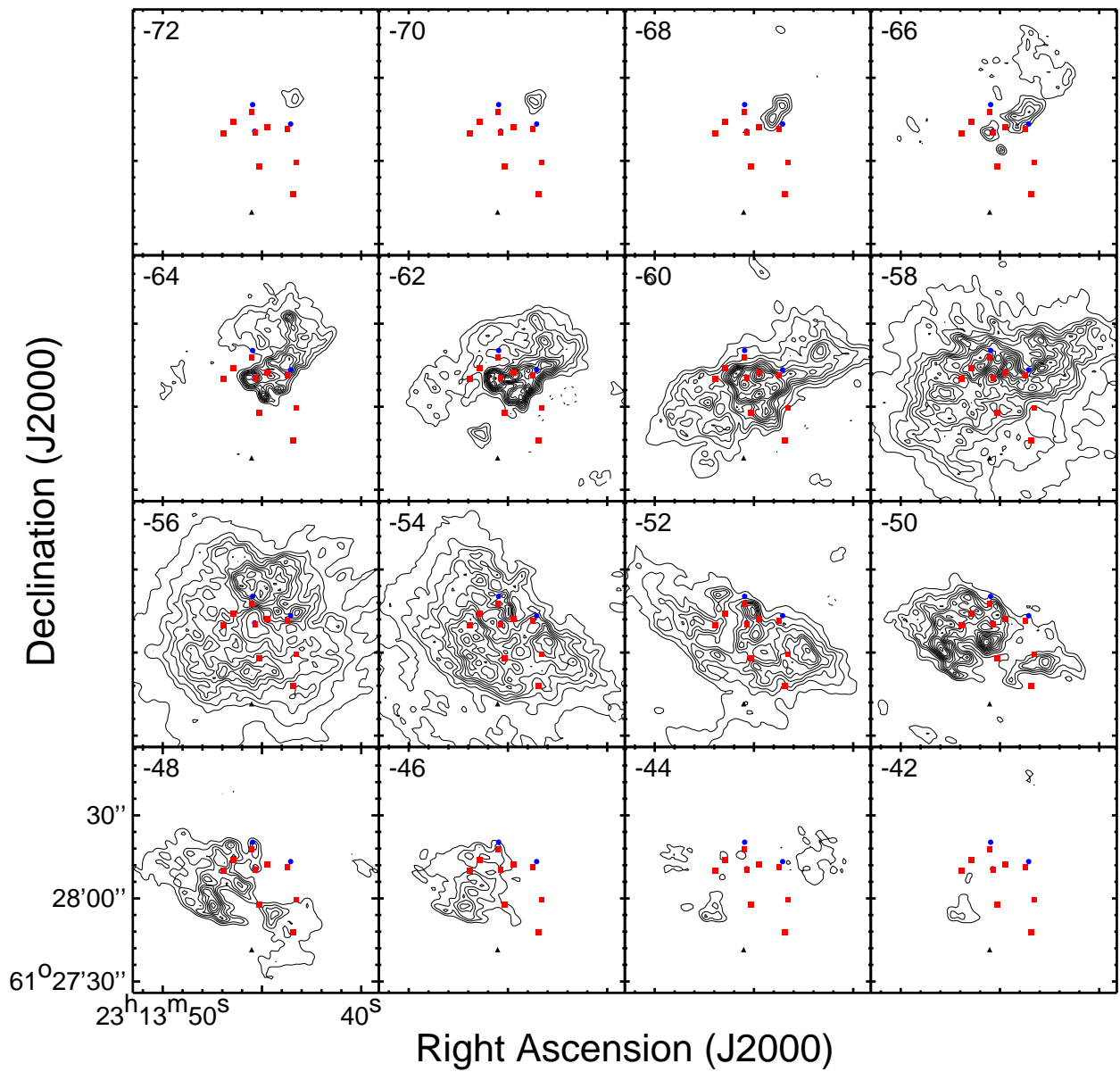} \caption{Velocity channel maps of
$^{13}$CO (2--1) emission obtained from the combined SMA and CSO
data. The starting and spacing contour levels are 0.8
Jy\,beam$^{-1}$ for channels from $-60$ to $-56$ km\,s$^{-1}$ and
0.45 Jy\,beam$^{-1}$ in outer line wings. Symbols are the same as
those in Figure \ref{co_chan}. \label{13co_chan}}
\end{figure}

The C$^{18}$O (2--1) emission is dominated by two dense clumps
residing on both sides of an absorption feature around MM1: the
western clump peaks at MM2 and shows extension from MM5 to MM4; the
eastern clump encompasses MM3 and MM6. While MM2 coincides with the
brightest C$^{18}$O peak, cores MM3--6 are more or less offset from
local C$^{18}$O peaks. Another interesting feature in C$^{18}$O is a
C-shaped shell seen in the southeast of the mm cores. In comparison
with the CO and $^{13}$CO (2--1) emission, the C$^{18}$O shell lies
immediately ahead of the brightest and furthest CO/$^{13}$CO clump
in the redshifted outflow (see Figure \ref{co_int} and Figure
\ref{13co_int}). This suggests that the C-shaped structure forms
from ambient gas which is compressed by the outflow from MM1.

Apart from emission and absorption toward MM1, the SO (5$_6$--4$_5$)
map shows bright emission around MM2, MM3, MM6, MM8, MM9, as well as
faint extensions toward MM4 and MM5. The SO map also shows clumps
without a counterpart in the dust continuum; for example, two bright
clumps to the southeast of MM6 are not seen in dust continuum or any
other line. It is unclear whether the SO emission in the two clumps
is associated with the MM6 outflow or arising from protostellar
cores whose dust continuum is below the detection limit of our SMA
observation.

Compared to C$^{18}$O (2--1) and SO (5$_6$--4$_5$), the CH$_3$OH
(8$_{-1,8}$--7$_{0,7}$) {\it E} line has a considerably higher
$E_{\mathrm {up}}$ (89 K) and appears to trace more compact
structures. In addition to MM1, cores MM3, MM6, MM7, MM8, and MM9
are clearly detected in this line. The CH$_3$OH peaks approximately
coincide with the dust continuum peaks, indicating the presence of
internal heating within each of the detected cores. Relatively faint
emission associated with MM5 is detected as well.

A serendipitous discovery from the CH$_3$OH (8$_{-1,8}$--7$_{0,7}$)
{\it E} map is a new 229.8 GHz maser at (R.A., decl.)$_{\mathrm
J2000}=(23^{\mathrm h}13^{\mathrm m}45.\!^{\mathrm s}52,
61^{\circ}27{'}41.\!{''}5)$, which is seen as unresolved and bright
emission with a peak flux of $\sim$1.64 Jy\,beam$^{-1}$ and a narrow
line width of less than 2 km\,s$^{-1}$ (Figure \ref{mom0}c and
Figure \ref{maser}). It is difficult to assess whether this maser
spot is associated with any of the mm continuum sources which are
more than 0.2 pc away. Another possibility is that the maser is
associated with a compact CH$_3$OH (8$_{-1,8}$--7$_{0,7}$) {\it E}
and SO (5$_6$--4$_5$) clump about 5$''$ to the north. The clump
probably traces a low- to intermediate-mass protostellar core whose
1.3 mm dust emission is below the detection limit of our SMA
observation. More interestingly, the maser spot lies right at the
tip of a compact CO outflow (Figure \ref{co_int}), suggesting that
the masing line is pumped by the outflow shocks, which is consistent
with the class I property of the transition. There has been very few
report on the detection of 229.8 GHz methanol maser
\citep{Slysh02,Qiu09a}, probably because of its high frequency hence
not covered by previous methanol maser surveys.

In the VLA NH$_3$ (1,1) and (2,2) data taken by Zheng et al. (2001),
emission in MM6--9 is clearly seen in both transitions (see Figure
\ref{nh3_mm69}), while an investigation of emission in MM2--5 is
severely affected by predominant absorption arising from MM1. As
there is no clear detection of a satellite line in Figure
\ref{nh3_mm69}, we assume the emission to be optically thin, and
deduce rotational temperatures, $T_R$, from the intensity ratio of
the two transitions following
$T_R=-41.7/ln(0.282\,T_{2,2;m}/T_{1,1;m}),$ where $T_{1,1;m}$ and
$T_{2,2;m}$ are the brightness temperatures of the main hyperfine
components of the NH$_3$ (1,1) and (2,2) transitions, respectively
\citep{Ho83}; the formula assumes equal line width for the two
transitions, which is probably a good approximation but difficult to
verify given the moderate sensitivity and spectral resolution. We
obtain $T_R$ of $29\pm13$, $34\pm12$, $33\pm14$, and $27\pm7$ K for
MM6--9, respectively; the uncertainties account for the $1\sigma$
noise of the spectra. Correcting for the depopulation effect
\citep{Danby88}, we estimate the kinetic temperatures from $T_R$ to
be 43, 58, 54, and 38 K for MM6--9, respectively.

\section{Discussion} \label{dis}

\subsection{Nature of the MM1 core}
The MM1 core has a size of about 0.05 pc, an H$_2$ number density of
order $10^7$ cm$^{-3}$, a characteristic temperature of 245 K, and
rich chemistry in complex organic molecules. All these indicate that
MM1 is a typical HMC. On the other hand, the presence of bright IR
and radio emission suggests that the core is at a relatively more
evolved stage compared to many other HMCs \citep[e.g.,][]{Chen06,
Zhang07, Girart09, Qiu09a}. The core also shows particularly rich
chemistry in large O-bearing molecules (e.g., CH$_3$OCHO and
C$_2$H$_5$OH), consistent with a relatively more evolved stage.

According to stellar structure models of Schaller et al. (1992), the
central young star embedded within MM1 has a mass of $\gtrsim25$
$M_{\odot}$. Redshifted absorption seen in several lines toward MM1
implies the presence of an accretion flow onto the central star or
star-disk system which may further increase the mass of the young O
star. The $\sim$20 $M_{\odot}$ core, however, is not equivalent to
the gas reservoir of the accretion. On the one hand, Myers (2009)
found that clustered and high-density regions form high-mass
protostars from both core and environment gas. There are also claims
that the accretion onto a forming O star may be fed by global infall
of the parent cloud \citep[e.g.,][]{Galvan09}. For NGC 7538 IRS 1,
it is unclear whether there is a global infall pertaining to the
accretion. Previous NH$_3$ (1,1), (2,2) observations revealed
\emph{blueshifted} absorption toward IRS 1 \citep[][also see Figure
\ref{absorption}]{Wilson83, Henkel84, Keto91}, which is attributed
to expanding or outflowing gas moving toward the observer. The
NH$_3$ absorption is optically thick ($\tau\gg1$, Henkel et al.
1984; Keto 1991), presumably tracing gas mostly residing in outer
layers surrounding the HMC. Thus the bulk of the outer gas appears
to be in an outward motion. On the other hand, various mechanisms,
e.g., outflows, heating and ionization, and accretion by nearby
protostars, can disperse dense gas and conspire to terminate the
accretion \citep{Myers09}. In particular, Keto (2007)
described a three-stage evolutionary sequence of an H{\scriptsize
II} region around a massive young star. In his model, the dynamics
and morphology of the ionized gas are determined by the ratio of the
ionization radius, $R_i$, and the gravitational radius, $R_b$,
defined by $GM_{\ast}/2c^2$, where $M_{\ast}$ is the mass of the
star and $c$ is the sound speed. With $M_{\ast}=25$ $M_{\odot}$ and
$c=10$ km\,s$^{-1}$ (for ionized gas temperature of $10^4$ K),
$R_{b}$ is 110 AU. The innermost ionized gas around IRS 1 has a
radius of $\sim$500 AU \citep{Campbell84a, Gaume95, Sandell09}. Thus, for
IRS 1, $R_i>R_b$, corresponding to the second to third stages in
Keto's model, that is, the ionized gas moves outward and
progressively expands toward the equatorial plane. In this picture
the accretion is confined to a narrow range of angle around the disk
and is close to termination.

\subsection{The MM1/IRS 1 outflow} \label{dis_outflow}
\begin{figure}
\epsscale{.9} \plotone{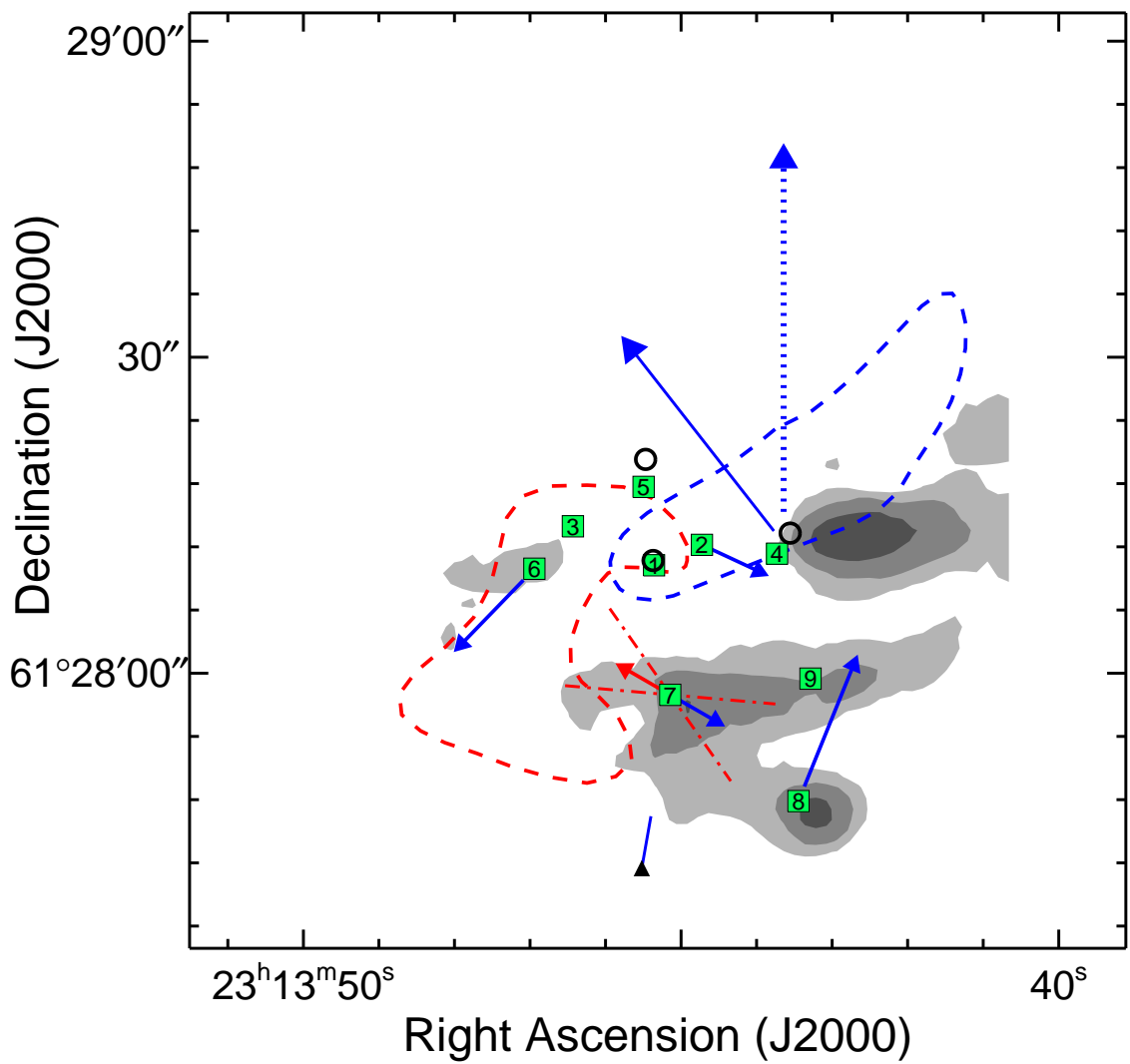} \caption{A schematic view of the
multiple outflows seen in the CO and $^{13}$CO emission. Dashed
curves roughly outline the MM1/IRS 1 outflow; arrows approximately
denote the orientation and extension of suggested outflows from the
other cores; two dash-doted lines bisecting at MM7 highlight a
bi-conical structure remarkable in $^{13}$CO; a PA$\sim$0$^{\circ}$
structure is show in a dotted arrow for its ambiguous interpretation
(see Section \ref{dis_outflow} for details). Blue and red colors
represent blue- and redshifted velocities of the structure,
respectively. Other symbols are the same as those in Figure
\ref{co_int}. The NH$_3$ (2,2) emission is shown in filled gray
contours (data from Zheng et al. 2001; see Section \ref{dis_outflow}
for discussions on a dense clump to the west of MM4).
\label{sketchy}}
\end{figure}

\begin{figure}
\epsscale{1} \plotone{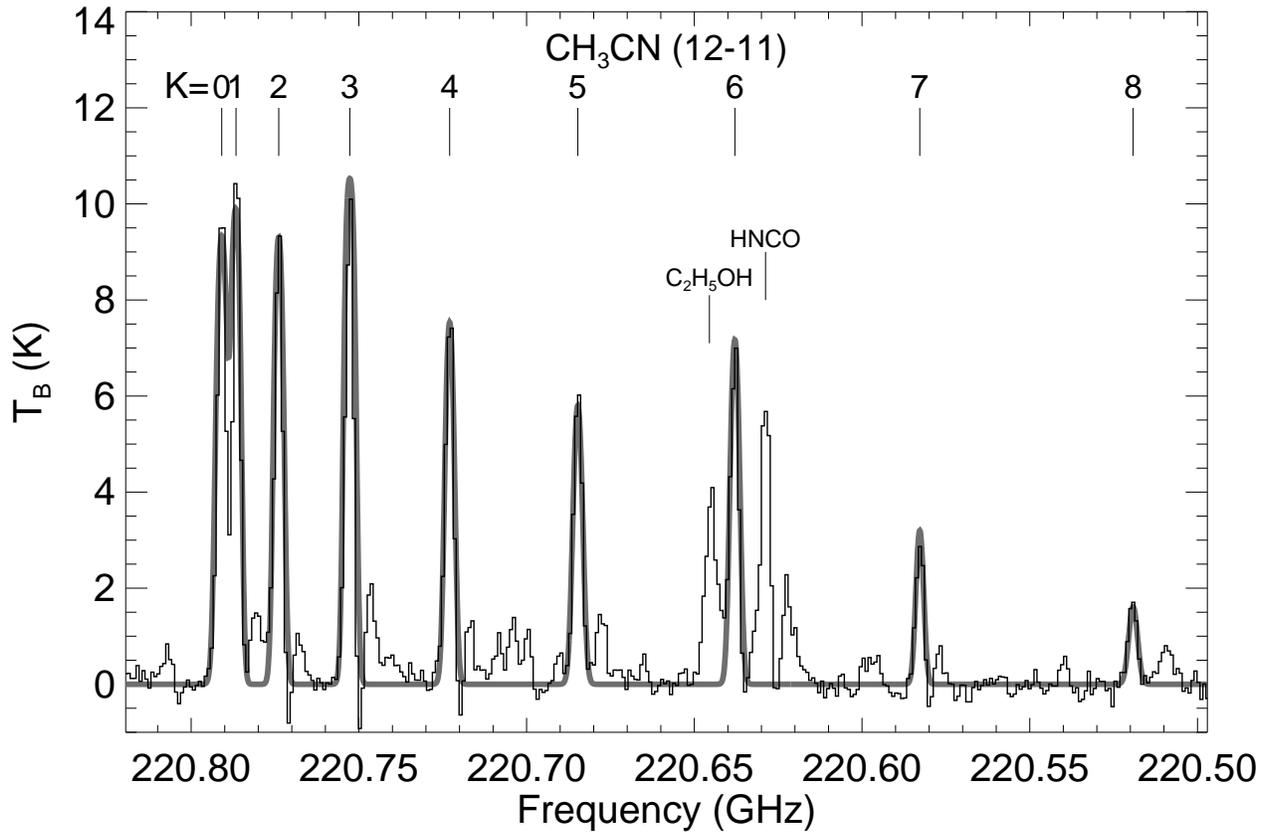} \caption{The observed spectra of
CH$_3$CN (12--11) toward MM1 in black histogram, overlaid with the
best fit LTE model in thick gray. \label{ch3cn}}
\end{figure}

\begin{figure}
\epsscale{.62} \plotone{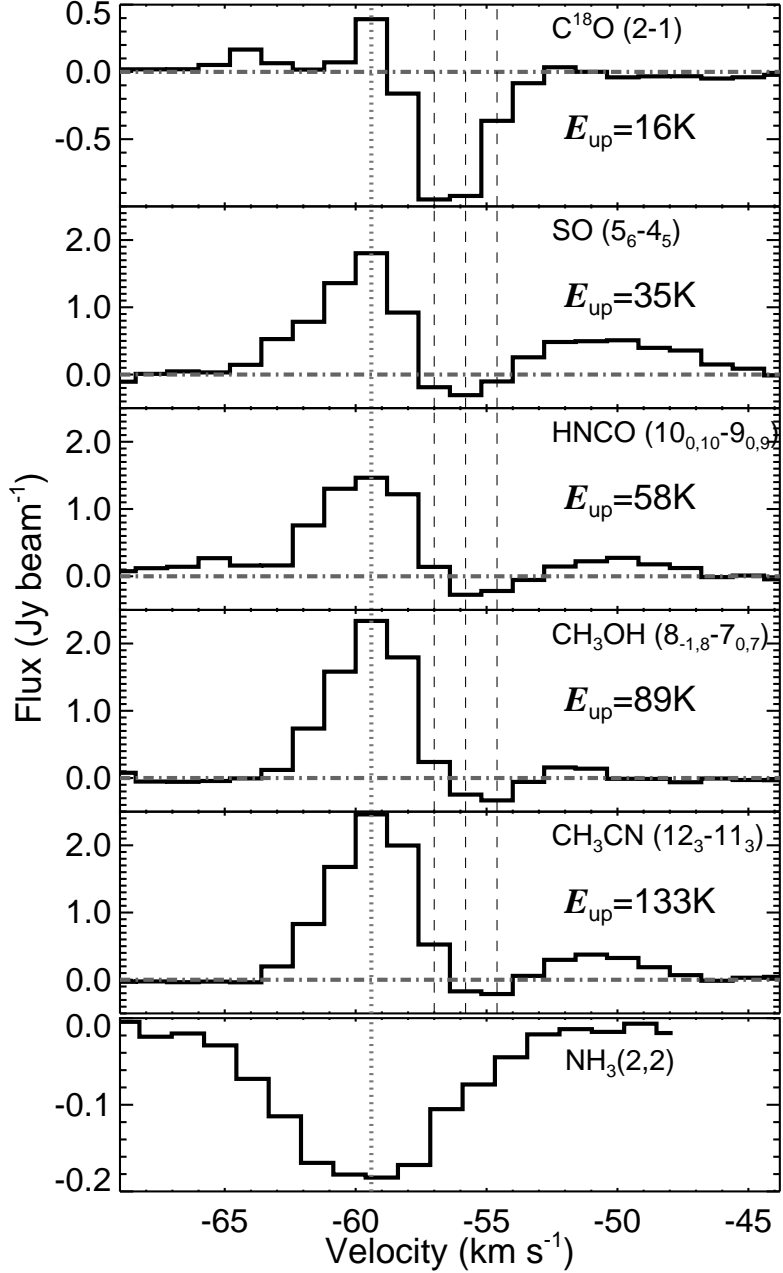} \caption{The spectra of C$^{18}$O
(2--1), SO ($5_6$--$4_5$), HNCO ($10_{0,10}$--$9_{0,9}$), CH$_3$OH
($8_{-1,8}$--$7_{0,7}$) {\it E}, and CH$_3$CN (12--11) $K$=3 toward
the peak of MM1. Three dashed lines denote channel velocities at
$-57$, $-55.8$, and $-54.6$ km\,s$^{-1}$, the peaking velocities of
the absorption. The dotted gray line denotes the channel of $-59.4$
km\,s$^{-1}$, the peaking velocity of the emission in all the lines.
The upper energy level of each transition is indicated in the
right-hand of each panel. For comparison, the absorption in NH$_3$
(2,2) is shown in the bottom panel (data from Zheng et al. 2001).
\label{absorption}}
\end{figure}

\begin{figure}
\epsscale{1} \plotone{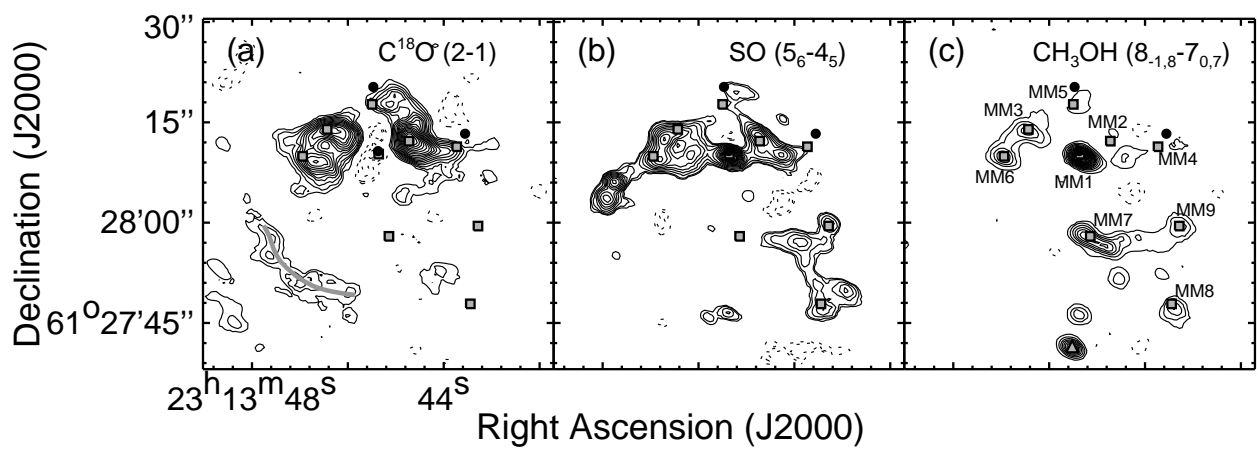} \caption{Moment-0 (velocity averaged)
maps shown in contours for the lines of (a) C$^{18}$O (2--1), (b) SO
($5_6$--$4_5$), (c) CH$_3$OH ($8_{-1,8}$--$7_{0,7}$) {\it E}.
Contour levels start from 50 mJy\,beam$^{-1}$ and increase with
steps of 25 mJy\,beam$^{-1}$ for the C$^{18}$O map, start from 48
mJy\,beam$^{-1}$ and increase with steps of 24 mJy\,beam$^{-1}$ for
the SO map, and start from 24 mJy\,beam$^{-1}$ and increase with
steps of 29 mJy\,beam$^{-1}$ for the CH$_3$OH map. Cores MM1--9 are
labeled as filled squares and for clarity, with their nomenclature
shown in the CH$_3$OH map; IRS 1--3 are denoted as three dots.
\label{mom0}}
\end{figure}

\begin{figure}
\epsscale{1} \plotone{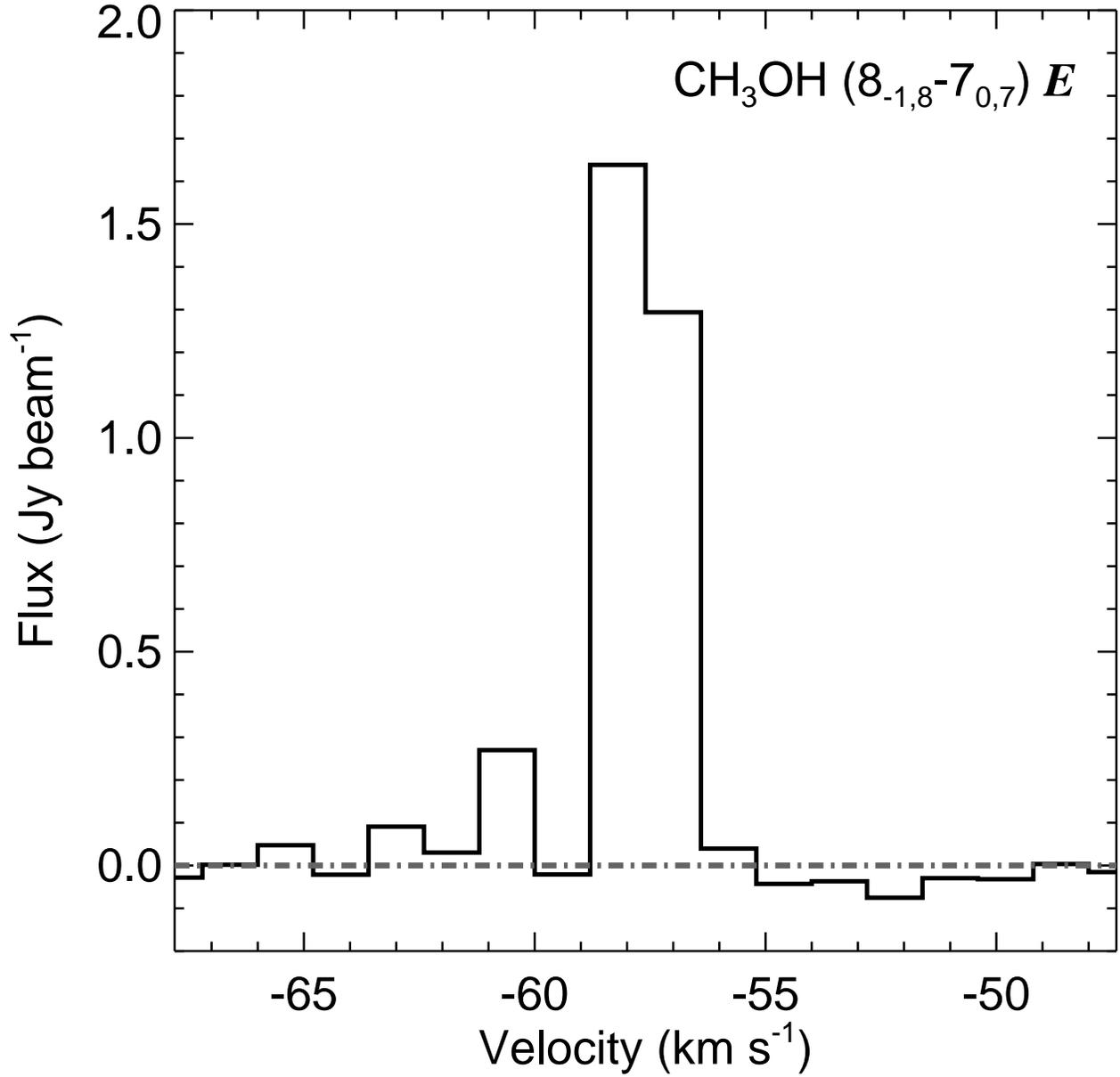} \caption{The velocity profile of the
229.8 GHz methanol maser discovered in Figure \ref{mom0}c (see
Section \ref{newmm_line} for details). \label{maser}}
\end{figure}

\begin{figure}
\epsscale{1} \plotone{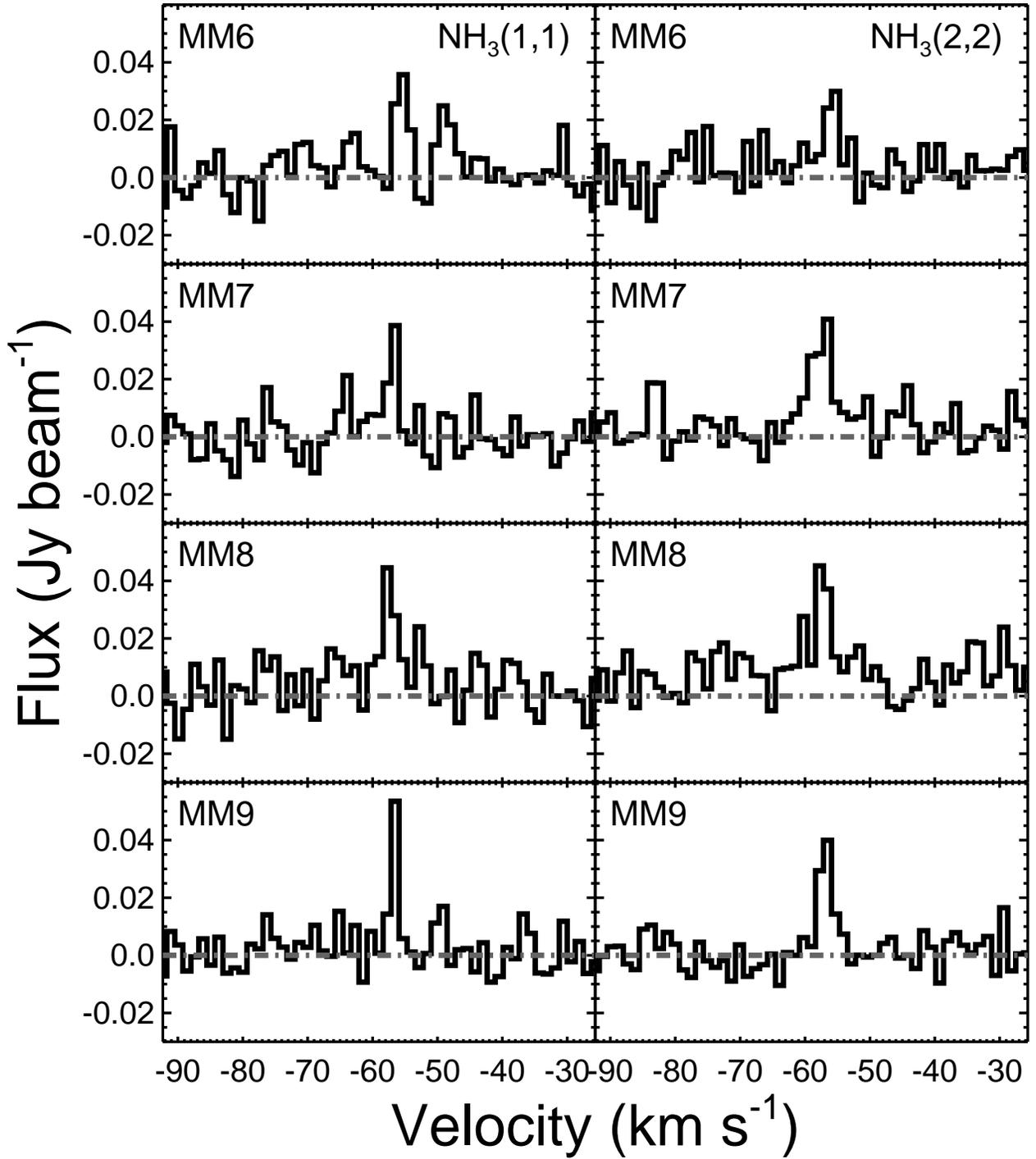} \caption{The NH$_3$ (1,1) and (2,2)
spectra in MM6--9; data taken from Zheng et al. (2001)
\label{nh3_mm69}}
\end{figure}

Multiple outflows are seen in the CO and $^{13}$CO maps. However,
the bulk outflowing gas is centered at the brightest core MM1 which
harbors the IRS 1 young O star. As outlined in Figure \ref{sketchy},
the blueshifted lobe of the MM1 outflow is dominated by a slightly
curved filament with a PA of about $-50^{\circ}$. In sub-arcsec
resolution observations, an elongated feature with an orientation
similar to that of the CO/$^{13}$CO filament is seen in radio
continuum \citep[$\lesssim1''$,][]{Campbell84a, Gaume95, Sandell04}
and mid-IR emission \citep[$\sim4''$,][]{DeBuizer05}. The 0.5 pc
structure seen in CO and $^{13}$CO (2--1) can be traced back to
about $0.\!''2$ ($\sim$500 AU) from the central source; within
$r\sim0.\!''2$, the ionized gas shows a double-lobed morphology in a
north-south orientation. In addition, near-IR observations toward
IRS 1 reveals a fan-shaped structure (opening angle
$\sim90^{\circ}$) extending from IRS 1 to the northwest
\citep[$\sim10''$,][]{Kraus06}. On a larger scale, the CO/$^{13}$CO
filament is seen projected against a dense clump revealed by the
NH$_3$ emission (see Figure \ref{sketchy}). The most likely scenario
capable of incorporating the inner ionized gas with a north-south
orientation and the northwestern structure extending from 500 AU to
0.5 pc from the central source seems to be a wide-angle wind from
IRS 1 which carves an outflow cavity. The axis of the wide-angle
wind is mostly along the north-south orientation and slightly
inclined to the west. The northwestern structure observed in radio,
mid-IR, and in CO and $^{13}$CO (2--1) emission traces the western
wall of the outflow cavity. The lack of appreciable emission from
the eastern wall is likely ascribed to the deficiency of dense gas
to the northeast of IRS 1 \citep[e.g.,][]{Zheng01}. Indeed, from a
close inspection of the ionized gas and hot dust immediately around
IRS 1, faint extension to the northeast does exist as well (Figure 2
in Campbell 1984; Figure 1b in Gaume et al. 1995; Figure 2b in De
Buizer \& Minier 2005). The CO/$^{13}$CO structure originating from
the vicinity of IRS 3 and MM4 and extending to the north
(PA$\sim$0$^{\circ}$) is puzzling; in Figure \ref{co_int} and Figure
\ref{13co_int}, the structure seems to be part of the MM1 outflow
but with an orientation very different from the inner part (i.e.,
the PA of the outflow varying from $-50^{\circ}$ to $0^{\circ}$). In
dust continuum or NH$_3$ maps of the region \citep{Zheng01,
Sandell04, Reid05, Pestalozzi06}, there appears to be no dense gas
immediately to the west of the structure, excluding the possibility
of being confined by ambient dense gas. From Figure \ref{sketchy},
the wide-angle wind seems to be expanding into an ambient dense
clump seen in NH$_3$ and as a consequence of this interaction, part
of the gas is deflected to the north, forming the
PA$\sim$0$^{\circ}$ structure. However, with the existing data it
cannot be ruled out that the structure is associated with IRS 3 or
MM4.

It is not straightforward either to unambiguously identify and
interpret the redshifted lobe of the MM1 outflow. In the wide-angle
wind scenario proposed above, the structure mostly consists of
ambient gas being entrained or swept up by the wide-angle wind and
tracing the eastern wall of the southern outflow cavity. The
slightly distorted appearance of the structure could be partly
caused by interaction with other outflows (e.g., the MM7 outflow).
An interesting feature is the C-shaped shell seen in C$^{18}$O. It
lies ahead of the furthest and brightest CO/$^{13}$CO clump and
features dense gas being compressed by part of the leading front of
the wide-angle wind. It is puzzling that the western wall of the
southern outflow cavity is unseen in CO or $^{13}$CO, as at least in
projection dense gas to the southwest of MM1 is seen in NH$_3$ and
dust continuum. One possibility is that in a three-dimensional
picture the southern lobe of the wide-angle wind is expanding into a
medium without dense gas in the west and cannot produce appreciable
emission.

Hollenbach et al. (1994) studied the photoevaporation of disks
around massive young stars and developed a model of the
photoevaporative wind from the disk. Lugo et al. (2004) performed a
parametric investigation of such a wind in NGC 7538 IRS 1 by
modeling the free-free emission as arising from the wind and found a
mass-loss rate of $1.1\times10^{-5}$ $M_{\odot}$\,yr$^{-1}$. As the
speed of the photoevaporative wind is $\sim$10--50 km\,s$^{-1}$
\citep{Hollenbach94}, comparable to that of CO outflows, the mass
loss rate of the wind seems too low to drive the NGC 7538 IRS 1
outflow, whose mass loss rate is of order $10^{-3}$
$M_{\odot}$\,yr$^{-1}$ (Table \ref{outflow_tbl}). The
photoevaporative wind may play a role in driving the IRS 1 molecular
outflow, however, the large amount of gas mass, momentum, and energy
in outflowing gas requires a more energetic wind as its driving
engine. In a \emph{momentum-driven} picture (working like a
snow-plow), a fast, $\sim$1000 km\,s$^{-1}$ wind with a mass loss
rate of $10^{-5}$ $M_{\odot}$\,yr$^{-1}$ is able to drive the
observed molecular outflow; in an \emph{energy-driven} situation
(working by ram pressure), the required mass loss rate is much
lower, i.e., $10^{-7}$ $M_{\odot}$\,yr$^{-1}$, provided the wind
speed reaches 1000 km\,s$^{-1}$. Stellar winds from O stars and
possible accretion-driven winds (e.g., a high-mass analogue of {\it
X}-winds in low-mass protostars) are expected to have velocities
reaching $\sim$1000 km\,s$^{-1}$, and energy-wise, are likely to
drive the massive and energetic molecular outflow.

\subsection{A massive cluster in the making}\label{cluster}
In single-dish submm and mm continuum maps of the NGC 7538 complex,
the most remarkable cloud has a size of about 1 pc and slightly
extends to the southeast \citep{Sandell04, Reid05, Pestalozzi06}.
Star formation activity within this cloud has long been thought to
be concentrated around IRS 1--3, and IRS 1 is the source that
attracted most of the interest. Our SMA 1.3 mm continuum observation
uncovers a total of nine cores with a spatial distribution roughly
following the shape of the cloud. Apart from MM1, which harbors the
IRS 1 young O star, cores MM2--9 are new detections. The detection
of these dusty cores is interesting in the context of cloud
fragmentation and cluster formation.

For MM1, with a number density of $1.7\times10^7$ cm$^{-3}$
(averaged within a beam and assuming a spherical geometry) and a
temperature of 245 K, one derives a thermal Jeans mass ($M_J$) of 17
$M_{\odot}$, which is comparable to the measured core mass and
compatible with the scenario that heating from the forming star may
have helped to suppress fragmentation of the envelope
\citep[e.g.,][]{Krumholz06, Krumholz08}. However, all the new cores
are very young, apparently lacking significant heating. These cores
are not detected in CH$_3$CN, and only part of them are seen in one
transition of CH$_3$OH (see Figure \ref{mom0}). For MM6--9, the
NH$_3$ data reveal temperatures of order 50 K (Section
\ref{newmm_line}). For MM2--5, a rough upper limit may be obtained
based on the absence of CH$_3$CN emission, which is a typical
thermometer of HMCs. For example, cores MM2--5 have masses close to
MM1; if the abundance ratio and spatial distribution of CH$_3$CN in
MM2--5 resembles that in MM1, at a temperature of 100 K, the
brightness temperature (averaged within a beam of $3''\times2''$) of
the CH$_3$CN (12--11) $K$=0--3 emission in MM2--5 would be about
2--4 K, which could have been detected at a $\gtrsim15\sigma$ level.
Alternatively, with the CH$_3$CN (12--11) $K$-ladder observed with
the SMA, Qiu \& Zhang (2009) derived a temperature of 110 K for a
HMC in the HH 80-81 massive star-forming region; if the HH 80-81 HMC
were at a distance of 2.65 kpc and observed with a $3''\times2''$
beam, the brightness temperature of the $K$=0,1 components would
have been around 1 K; cores MM2--5 in NGC 7538 have masses
comparable to that of the HH 80--81 HMC, while the CH$_3$CN emission
is not detected ($\sigma\sim0.13$ K). All this indicates that MM2--5
have temperatures well below 100 K. Sandell \& Sievers (2004) found
a dust temperature of 40 K for submm condensations around IRS 1.
This probably provides a rough estimate of the averaged temperature
of MM2--9, as part of them roughly coincide with the submm
condensations.

\begin{figure}
\epsscale{1} \plotone{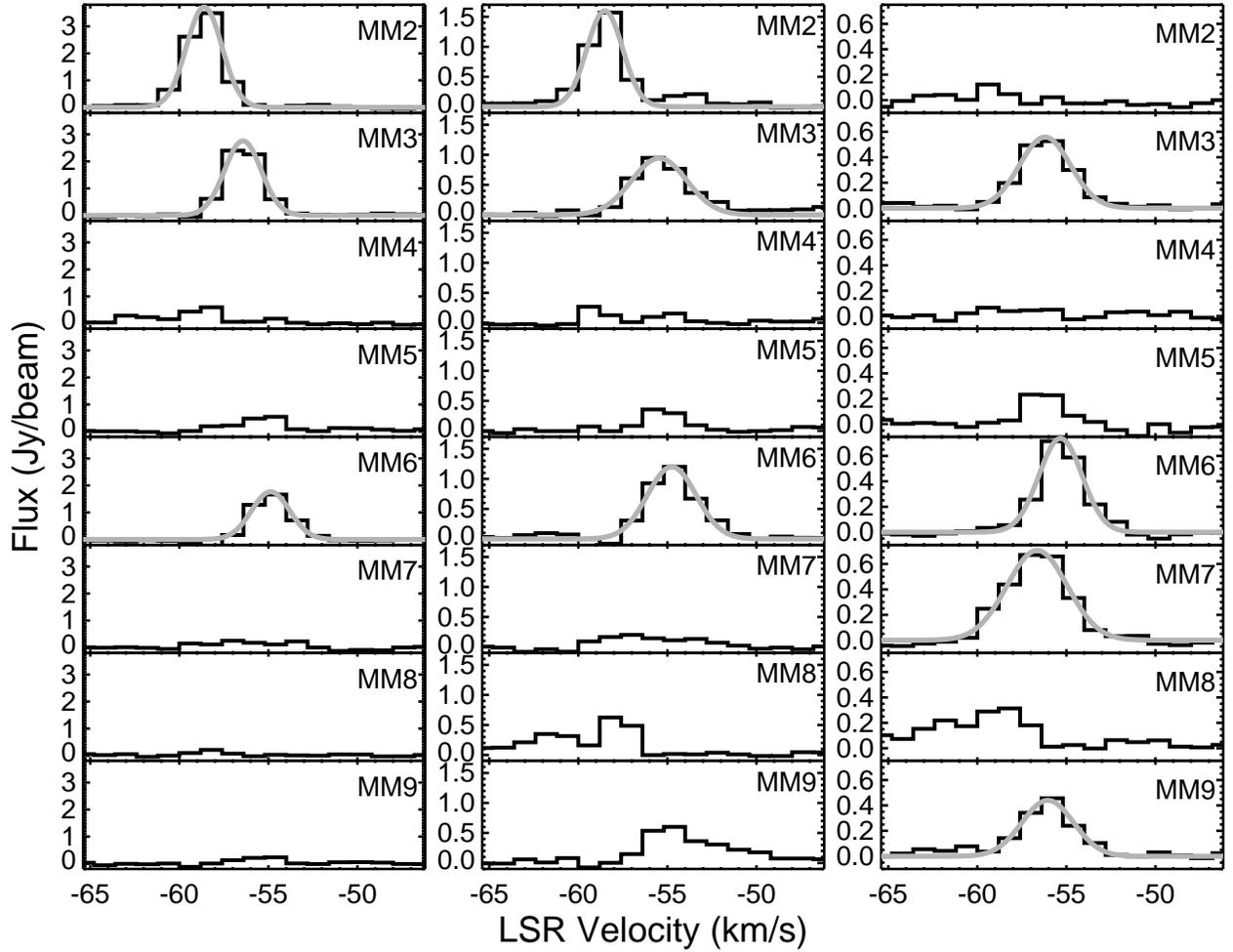} \caption{The C$^{18}$O (2--1) {\it
Left}), SO (5$_6$--4$_5$) ({\it Middle}), and CH$_3$OH
(8$_{-1,8}$--7$_{0,7}$) {\it E} ({\it Right}) spectra toward the
peaks of cores MM2--9 shown in histogram, with Gaussian fittings
overlaid in thick gray lines. \label{linwid}}
\end{figure}

MM2--9 have number densities of $\sim0.4$--$1.2\times10^7$
cm$^{-3}$, or mass column densities $\gtrsim1$ g\,cm$^{-2}$,
matching a proposed threshold of 1 g\,cm$^{-2}$ for molecular clouds
capable of avoiding fragmentation and forming massive stars
\citep{Krumholz08}. But the masses of the new cores are well above
$M_J$, implying the importance of supersonic turbulence and/or
magnetic field in cloud fragmentation and possible star formation
within the new cores \citep[e.g.,][]{Zhang09}. While an
investigation of the magnetic field awaits new observations, a
glimpse of turbulence can be obtained by looking into the emission
lines arising from these cores. Figure \ref{linwid} shows spectra in
C$^{18}$O (2--1), SO (5$_6$--4$_5$), and CH$_3$OH
(8$_{-1,8}$--7$_{0,7}$) {\it E} at the peaks of MM2--9. A gaussian
fitting is performed on lines with sufficient signal-to-noise
ratios. The derived line widths of $\sim$2.5--4 km\,s$^{-1}$ are
more than 5 times larger than the thermal line width. This strongly
suggests the presence of supersonic turbulence which may play an
important role in suppressing fragmentation, if the line widths are
not dominated by kinematics related to star formation processes
(e.g., outflow, infall, and rotation).

Information on potential ongoing star formation within the new cores
can be obtained from their associations with CO outflows, H$_2$O
masers, and warm and dense gas condensations. Cores MM2, MM3, MM4,
and MM7 are all associated with H$_2$O maser emission (Figure
\ref{cont_map}), indicative of active star formation. Cores MM2,
MM6, MM7, MM8, and probably MM4 as well, are associated with CO
outflows (Figure \ref{sketchy} and Section \ref{outflow}), implying
the presence of protostars within these cores. MM9 shows the
faintest dust continuum; although lacking apparent association with
an outflow or H$_2$O maser spot, it is very likely internally heated
by an intermediate-mass protostar as it shows centrally peaked
emission in CH$_3$OH and SO (Figure \ref{mom0}). There is no clear
signature of star formation toward MM5, which also shows a
relatively flattened profile in dust emission. It is more likely a
gas and dust clump externally heated by IRS 2 \citep[probably
through a stellar-wind bow shock,][]{Bloomer98}, rather than an
internally heated star-forming core. In short, seven out of the
eight new cores appear to be forming stars. The star-forming cores
have masses of $\sim5$--24 $M_{\odot}$ and high densities, and most
of them show bright, centrally peaked CH$_3$OH emission. Although
CH$_3$OH lines are seen in hot corinos \citep[warm and dense
envelopes around low-mass protostars,][]{Ceccarelli07}, the emission
could not be detected if the hot corinos were at distances $>2$ kpc
as compared to the case of IRAS 16293-2422
\citep{vanDishoeck95,Kuan04,Chandler05}. Thus, the new cores seen in
CH$_3$OH are expected to be forming intermediate- to high-mass
stars. MM4 is not detected in CH$_3$OH and its immediate northwest
is deficient of emission in C$^{18}$O and SO; molecular line
emission from this core is probably hampered by UV radiation from
IRS 3, which has begun to ionize the surrounding and is not heavily
obscured by a dusty cocoon. MM2 is the most massive of the new cores
and coincides with the brightest C$^{18}$O peak, but is not seen in
CH$_3$OH. With a relatively low $E_{\mathrm {up}}$ (16 K), the
C$^{18}$O (2--1) emission preferentially traces column density
enhancements rather than temperature increase
\citep[e.g.,][]{Wyrowski99}. The CH$_3$OH (8$_{-1,8}$--7$_0,7$) $E$
emission (with a much higher $E_{\mathrm {up}}$ of 89 K) is more
sensitive to the heating. Therefore, MM2 is presumably the youngest
of the new star-forming cores.

The NGC 7538 molecular complex has been thought to exhibit
sequential massive star formation propagating from northwest to
southeast \citep{Elmegreen77, Werner79}. In this picture, the IRS
1--3 region lies ahead of the ionization front of the optical
H{\scriptsize II} region and represents an intermediate evolutionary
phase between the ionizing stars of the H{\scriptsize II} region in
the northwest and deeply embedded massive protostars in the
southeast \citep[i.e., NGC 7538 S and IRS 9,][]{Werner79}. By
zooming in the parent cloud of IRS 1--3, our SMA observations
provide new insights into this scenario. With the newly discovered
star-forming cores, the IRS 1 young O star is not only accompanied
by more evolved sources IRS 2 and IRS 3, but also surrounded by a
group of intermediate- to high-mass protostars distributed over the
parent cloud. There is no pronounced propagation of star formation
across the cloud. In comparison with massive protostars in IRS 9 or
NGC 7538 S, IRS 1--3 are certainly more evolved, but all the newly
discovered star-forming cores are at very young evolutionary stages.
In particular MM2, which is the closest to MM1, is apparently
younger than IRS 9, which is visible in the near- to mid-IR
\citep{Werner79}, and even younger than NGC 7538 S, which shows
bright CH$_3$CN emission \citep{Sandell10}. The pc-sized cloud
appears to be in an active phase of cluster formation, with around
10 OB-type stars and probably orders of magnitude more low-mass
stars according to a typical IMF \citep[initial mass function,
e.g.,][]{Salpeter55,Kroupa02}. Indeed, within the volume occupied by
high- and intermediate-mass protostars described above, Kraus et al.
(2006) in their 2.2 $\mu$m speckle images found a number of weak,
compact sources. It is interesting to speculate that these are
already more evolved lower mass members of the same forming star
cluster of which IRS 1--3 and the new protostars (MM2--4 and MM6--9)
define the higher mass end of the mass distribution. Here, a
comparison with the well-studied Orion Nebular Cluster (ONC) is
instructive. As argued by Palla \& Stahler (1999) and Huff \&
Stahler (2006), the highest mass ONC member stars comprising the
``Trapezium'' at the center of the cluster are $\sim10^5$ yr old,
while the bulk of the lower mass cluster members formed 1--3 million
years ago. In fact, one may consider IRS 1--3, MM2--4 and MM6 to be
a Trapezium system in formation. The maximal physical separation
between these objects, 0.30 pc, is much larger than the maximal
separation between members of the ONC Trapezium, 0.047 pc, i.e.,
between $\theta^1$~Ori~D and $\theta^1$~Ori~E, assuming a distance
of 417 pc to the ONC \citep{Menten07}. It is, however, well within
the range of other Trapezium-type systems. A Trapezium-type system,
as defined by Ambartsumian (1954; 1955; 1958), of which
$\theta^1$~Ori is the prototype, is a multiple star whose
separations are all roughly of the same order of magnitude. While
their cosmogonic role is somewhat unclear, it is a well-established
fact that such systems are frequently found as the highest mass
systems at the centers of clusters of lower mass stars
\citep{Sharpless54}. A catalog of 87 Trapezia with maximum
separations, $r_{\rm max}$, within 0.2 pc was published by
Salukvadze (1978), while Salukvadze \& Dzhavakhishvili (1988)
reported 15 wider systems with $r_{\rm max}$ between 1.1 and 5.7 pc.
Thus the dimensions of the nascent NGC 7538 IRS 1 Trapezium are well
comparable to values generally found for such systems.

\section{Summary} \label{conclu}
We present a 2--3$''$ resolution study in the 1.3 mm waveband of the
well-known massive star-forming region NGC 7538 IRS 1--3. The gas
and dust core MM1, which harbors the IRS 1 young O star, is massive
($\sim20$ $M_{\odot}$), hot ($\sim245$ K), and shows bright line
emission in complex organic molecules. Redshifted absorption seen in
several high-density tracing lines indicates that the central young
O star or star-disk system is still accreting gas from MM1. In
comparison with a theoretical model on early evolution of
H{\scriptsize II} regions \citep{Keto07}, the accretion seems to be
confined to a narrow range of angle about the equatorial plane.

Neither of the two nearby sources IRS 2 and IRS 3 is associated with
a dense dusty envelope, confirming that the two luminous sources are
at most moderately obscured.

Sensitive and high-angular-resolution observations also reveal eight
new cores MM2--9 embedded within the parent cloud of IRS 1--3. These
cores have masses much larger than the thermal Jeans mass,
indicating the importance of turbulence and/or magnetic field in
cloud fragmentation. The cores are seen in one or more of the
C$^{18}$O (2--1), SO ($5_6$--$4_5$), and CH$_3$OH
($8_{-1,8}$--$7_{0,7}$) {\it E} lines; the line widths are
significantly larger than the thermal line width, attributed to
supersonic turbulence and/or kinematics related to ongoing star
formation. Seven out of the eight cores appear to be forming
intermediate- to high-mass stars, of which the most massive ones
appear to be members of a nascent Trapezium system. Characterized by
the deeply embedded young O star IRS 1, two moderately obscured
sources IRS 2 and IRS 3, and seven new intermediate- to high-mass
protostellar cores, and taking into account presumably more unseen
low-mass (proto)stars, a massive cluster is forming from a pc-sized
cloud exposed to the ionization front of an optical H{\scriptsize
II} region.

The CO and $^{13}$CO maps obtained from the combined SMA and
single-dish data reveals an outflow scenario far more complicated
than a bipolar outflow seen in single-dish observations. Multiple
outflows are seen, and the MM1/IRS 1 outflow is dominating the mass
of 50 $M_{\odot}$, momentum of $4.6\times10^2$ $M_{\odot}$
km\,s$^{-1}$, and energy of $4.9\times10^{46}$ erg calculated for
the gas moving at velocities of $\leq-64$ km\,s$^{-1}$ and $\geq-48$
km\,s$^{-1}$. Comparing our CO, $^{13}$CO maps with the existing
radio and IR observations we suggest that the MM1/IRS 1 molecular
outflow is likely driven by a wide-angle wind from the forming O
star or star-disk system, and the wind is expanding into a medium
with inhomogeneous density distribution, creating asymmetric and
curved structures seen in CO/$^{13}$CO.

\acknowledgments We are grateful to Dr. Jingwen Wu and Dr. Ruisheng
Peng for help in obtaining the $^{13}$CO (2--1) observations at the
CSO. We acknowledge the anonymous referee for construtive comments
which improved the paper.

\clearpage

\clearpage

\clearpage

\clearpage

\clearpage

\clearpage

\clearpage

\clearpage

\clearpage

\clearpage

\clearpage

\clearpage

\clearpage

\begin{deluxetable}{cccllr}
\tablewidth{0pc} \tablecaption{Measured parameters and masses of mm
continuum cores \label{mm_tbl}}
\tablehead{ \colhead{} & \colhead{R.A.} & \colhead{Decl.} & \colhead{Peak Flux} & \colhead{Total Flux} & \colhead{Mass\tablenotemark{a}} \\
\colhead{} & \colhead{(J2000)} & \colhead{(J2000)} &
\colhead{(Jy\,beam$^{-1}$)} & \colhead{(Jy)} &
\colhead{($M_{\odot}$)} }

\startdata  MM1 & $23^{\mathrm h}13^{\mathrm m}45.358^{\mathrm s}$ & $61^{\circ}28'10.25''$ &  3.3  & 3.6   & $20.1\pm11.6$ \\
            MM2 & $23^{\mathrm h}13^{\mathrm m}44.724^{\mathrm s}$ & $61^{\circ}28'12.19''$ &  0.18 & 0.37  & $24.4\pm12.7$ \\
            MM3 & $23^{\mathrm h}13^{\mathrm m}46.432^{\mathrm s}$ & $61^{\circ}28'13.95''$ &  0.17 & 0.31  & $20.5\pm10.7$ \\
            MM4 & $23^{\mathrm h}13^{\mathrm m}43.729^{\mathrm s}$ & $61^{\circ}28'11.35''$ &  0.11 & 0.22  & $14.5\pm7.8$ \\
            MM5 & $23^{\mathrm h}13^{\mathrm m}45.496^{\mathrm s}$ & $61^{\circ}28'17.69''$ &  0.11 & 0.30  & $19.8\pm10.6$ \\
            MM6 & $23^{\mathrm h}13^{\mathrm m}46.942^{\mathrm s}$ & $61^{\circ}28'09.92''$ &  0.10 & 0.15  & $9.1\pm5.9$ \\
            MM7 & $23^{\mathrm h}13^{\mathrm m}45.143^{\mathrm s}$ & $61^{\circ}27'57.95''$ & 0.082 & 0.12  & $5.2\pm2.9$ \\
            MM8 & $23^{\mathrm h}13^{\mathrm m}43.446^{\mathrm s}$ & $61^{\circ}27'47.86''$ & 0.067 & 0.10  & $4.7\pm3.0$ \\
            MM9 & $23^{\mathrm h}13^{\mathrm m}43.286^{\mathrm s}$ & $61^{\circ}27'59.49''$ & 0.049 & 0.088 & $6.2\pm3.0$ \\
\enddata
\tablenotetext{a}{Calculated with a dust opacity $\kappa_{\mathrm
{225GHz}}=1$ cm$^2$\,g$^{-1}$, and temperatures of 245 K for MM1, 40
K for MM2--5, and 43, 58, 54, 38 K for MM6--9; see the discussion in
Section \ref{cont} for details about the mass uncertainties.}
\end{deluxetable}

\clearpage

\begin{deluxetable}{cccccc}
\tablewidth{0pc} \tablecaption{Derived outflow parameters \label{outflow_tbl}}
\tablehead{ \colhead{$M_{\mathrm {out}}$\tablenotemark{a}} & \colhead{$P_{\mathrm {out}}$\tablenotemark{b}} & \colhead{$E_{\mathrm {out}}$} & \colhead{$t_{\mathrm {dyn}}$} & \colhead{$\dot{M}_{\mathrm {out}}$} & \colhead{$\dot{P}_{\mathrm {out}}$} \\
\colhead{($M_{\odot}$)} & \colhead{($M_{\odot}$ km\,s$^{-1}$)} &
\colhead{(erg)} & \colhead{(yr)} & \colhead{($M_{\odot}$ yr$^{-1}$)}
& \colhead{($M_{\odot}$ km\,s$^{-1}$ yr$^{-1}$}) }

\startdata
           50 & $4.6\times10^2$ & $4.9\times10^{46}$ & $2\times10^4$ & $ 2.5\times10^{-3}$ & $2.3\times10^{-2}$ \\
\enddata
\tablenotetext{a}{Calculated for outflowing gas at $\leq-64$
km\,s$^{-1}$ and $\geq-48$ km\,s$^{-1}$;} \tablenotetext{b}{Without
correcting for an unknown inclination angle; this affects all the
parameters except $M_{\mathrm {out}}$;}
\end{deluxetable}

\clearpage

\begin{deluxetable}{llrr}
\tablewidth{0pc} \tablecaption{Molecular lines detected in
MM1\tablenotemark{a} \label{line_tbl}}
\tablehead{ \colhead{Frequency\tablenotemark{b}} & \colhead{Line\tablenotemark{c}} & \colhead{$E_{\mathrm {up}}$} & \colhead{Peak\tablenotemark{d}}\\
\colhead{(GHz)} & \colhead{} & \colhead{(K)} & \colhead{(Jy\,beam$^{-1}$)} }

\startdata
           219.560357 & C$^{18}$O 2--1 & 16 & 0.52 \\
           219.949442 & SO $5_6$--$4_5$ & 35 &  2.45 \\
           229.347628 & SO$_2$ $11_{5,7}$--$12_{4,8}$ & 122 & 0.29 \\
           231.060983 & OCS 19--18 & 111 & 2.55 \\
           219.908525 & H$_2^{13}$CO $3_{1,2}$--$2_{1,1}$ & 33 & 0.81 \\
           219.656710 & HNCO $10_{3,8(7)}$--$9_{3,7(6)}$ & 448 & 0.56 \\
           219.733850 & HNCO $10_{2,11(10)(9)}$--$9_{2,10(9)(8)}$ & 231 & 1.33 \\
           219.798320 & HNCO $10_{0,11(10)(9)}$--$9_{0,10(9)(8)}$ & 58 & 1.53 \\
           220.585200 & HNCO $10_{1,11(10)(9)}$--$9_{1,10(9)(8)}$ & 102 & 1.46 \\
           220.038072 & HCOOH $10_{0,10}$--$9_{0,9}$ & 59 & 0.66 \\
           220.178196 & CH$_2$CO $11_{1,11}$--$10_{0,10}$ & 77 & 0.98 \\
           219.983675 & CH$_3$OH $25_{3,22}$--$24_{4,20} E$ & 803 & 0.75 \\
           219.993658 & CH$_3$OH $23_{5,19}$--$22_{6,17} E$ & 777 & 0.64 \\
           220.078561 & CH$_3$OH $8_{0,8}$--$7_{1,6} E$ & 97 & 2.92 \\
           220.886784 & CH$_3$OH $31_{-2}$--$31_{1}$ & 1182 & 0.45 \\
           229.589056 & CH$_3$OH $15_{4,11}$--$16_{3,13} E$ & 375 & 4.59 \\
           229.758756 & CH$_3$OH $8_{-1,8}$--$7_{0,7} E$ & 89 & 2.86 \\
           229.864121 & CH$_3$OH $19_{5,15}$--$20_{4,16} A+$ & 579 & 1.79 \\
           229.939095 & CH$_3$OH $19_{5,14}$--$20_{4,17} A-$ & 579 & 1.93 \\
           230.027047\tablenotemark{e} & CH$_3$OH $3_{-2,2}$--$4_{-1,4} E$ & 40 & 3.36 \\
           230.368763\tablenotemark{f} & CH$_3$OH $22_{4,18}$--$21_{5,17} E$ & 684 & 1.56 \\
           220.362774\tablenotemark{g} & $^{13}$CH$_3$OH $17_{-3,15}$--$18_{0,18}$ & 410 & 0.41 \\
           220.475807 & CH$_3$CN $12_8$--$11_8$ & 526 & 0.39 \\
           220.539324 & CH$_3$CN $12_7$--$11_7$ & 419 & 0.71 \\
           220.594423 & CH$_3$CN $12_6$--$11_6$ & 326 & 1.87 \\
           220.641084 & CH$_3$CN $12_5$--$11_5$ & 248 & 1.66 \\
           220.679287 & CH$_3$CN $12_4$--$11_4$ & 183 & 2.07 \\
           220.709017 & CH$_3$CN $12_3$--$11_3$ & 133 & 2.98 \\
           220.730261 & CH$_3$CN $12_2$--$11_2$ &  98 & 2.79 \\
           220.743011 & CH$_3$CN $12_1$--$11_1$ &  76 & 3.05 \\
           220.747261 & CH$_3$CN $12_0$--$11_0$ &  69 & 2.78 \\
           219.268715 & CH$_3$CHO $31_{5,27}$--$32_{2,30}$ & 517 & 0.43 \\
           230.301880 & CH$_3$CHO $12_{2,11}$--$11_{2,10}$ & 81 & 0.54 \\
           219.411703 & CH$_3$OCHO $18_{10,8}$--$17_{10,7}$ & 355 & 0.39 \\
           219.763947 & CH$_3$OCHO $18_{9,9}$--$17_{9,8}$ & 342 & 0.64 \\
           219.822126 & CH$_3$OCHO $18_{10,9(8)}$--$17_{10,8(7)}$ & 355 & 0.41 \\
           220.030339 & CH$_3$OCHO $18_{9,10}$--$17_{9,9}$ & 342 & 0.57 \\
           220.166888 & CH$_3$OCHO $17_{4,13}$--$16_{4,12}$ & 103 & 1.26 \\
           220.190285 & CH$_3$OCHO $17_{4,13}$--$16_{4,12}$ & 103 & 1.32 \\
           220.258096 & CH$_3$OCHO $18_{8,10}$--$17_{8,9}$ & 331 & 0.41 \\
           220.369877 & CH$_3$OCHO $18_{8,10}$--$17_{8,9}$ & 331 & 0.52 \\
           220.913955 & CH$_3$OCHO $18_{7,12}$--$17_{7,11}$ & 321 & 0.36 \\
           220.946352 & CH$_3$OCHO $18_{7,11}$--$17_{7,10}$ & 321 & 0.34 \\
           220.977984 & CH$_3$OCHO $18_{15,3}$--$17_{15,2}$ & 250 & 0.38 \\
           220.985330 & CH$_3$OCHO $18_{7,11}$--$17_{7,10}$ & 321 & 0.45 \\
           220.998335 & CH$_3$OCHO $18_{15,4}$--$17_{15,3}$ & 250 & 0.52 \\
           221.047791 & CH$_3$OCHO $18_{14,5(4)}$--$17_{14,4(3)}$ & 231 & 0.50 \\
           221.066933 & CH$_3$OCHO $18_{14,5}$--$17_{14,4}$ & 231 & 0.36 \\
           221.141129 & CH$_3$OCHO $18_{13,6(5)}$--$17_{13,5(4)}$ & 213 & 0.97 \\
           229.405021 & CH$_3$OCHO $18_{3,15}$--$17_{3,14}$ & 111 & 1.44 \\
           229.420342 & CH$_3$OCHO $18_{3,15}$--$17_{3,14}$ & 111 & 1.45 \\
           230.293951 & CH$_3$OCHO $22_{9,13}$--$22_{8,14}$ & 204 & 0.75 \\
           230.315800\tablenotemark{h} & CH$_3$OCHO $22_{9,14}$--$22_{8,15}$ & 204 & 0.69 \\
           230.878810 & CH$_3$OCHO $18_{4,14}$--$17_{14,13}$ & 301 & 0.45 \\
           220.601927 & C$_2$H$_5$OH $13_{1,13}$--$12_{0,12}$ & 75 & 1.08 \\
           229.491131\tablenotemark{i} & C$_2$H$_5$OH $17_{5,12}$--$17_{4,13}$ & 160 & 0.66 \\
           229.818039 & C$_2$H$_5$OH $26_{5,22}$--$26_{4,23}$ & 385 & 0.43 \\
           230.177998 & C$_2$H$_5$OH $45_{2,44}$--$44_{3,43}$ & 894 & 0.41 \\
           230.230743 & C$_2$H$_5$OH $13_{2,11}$--$12_{2,10}$ & 143 & 0.66 \\
           230.473005\tablenotemark{j} & C$_2$H$_5$OH $18_{3,16}$--$17_{4,14}$ & 215 & 0.38 \\
           230.672554 & C$_2$H$_5$OH $13_{2,11}$--$12_{2,10}$ & 139 & 0.66 \\
           230.793764\tablenotemark{k} & C$_2$H$_5$OH $6_{5,1}$--$5_{4,1}$ & 105 & 0.53 \\
           230.953778 & C$_2$H$_5$OH $16_{5,11}$--$16_{4,12}$ & 146 & 0.75 \\
           230.991374 & C$_2$H$_5$OH $14_{0,14}$--$13_{1,13}$ & 86 & 1.08 \\
           220.893110 & CH$_3$OCH$_3$ $23_{4,20}$--$23_{3,21}$ & 275 & 0.40 \\
           230.141374 \tablenotemark{l}& CH$_3$OCH$_3$ $25_{4,22}$--$25_{3,22}$ & 319 & 0.51 \\
           229.234\tablenotemark{m} & --- & --- & 0.29 \\
           230.578\tablenotemark{m} & --- & --- & 0.65 \\
\enddata
\tablenotetext{a}{We list lines (excluding CO and $^{13}$CO 2--1)
detected above 6$\sigma$ (0.21 Jy\,beam$^{-1}$);}
\tablenotetext{b}{Rest frequencies obtained from the JPL catalog at
http://spec.jpl.nasa.gov/ftp/pub/catalog/catform.html;}
\tablenotetext{c}{Quantum numbers in parentheses denote degeneracy,
for instance, HNCO $10_{3,8(7)}$--$9_{3,7(6)}$ represents HNCO
$10_{3,8}$--$9_{3,7}$ and HNCO $10_{3,7}$--$9_{3,6}$;}
\tablenotetext{d}{Derived with the MIRIAD task MAXFIT;}
\tablenotetext{e}{Under LTE, the peak flux is unexpectedly higher
than other CH$_3$OH lines (e.g., CH$_3$OH $(8_{-1,8}$--$7_{0,7})
E$), which may be partly due to blending with C$_3$H$_7$CN
($52_{6,46}$--$51_{6,45}$), $\nu_{\circ}$=230.027116 GHz and
C$_2$H$_3$CN ($4_{3,1}$--$4_{2,2}$), $\nu_{\circ}$=230.026216 GHz;}
\tablenotetext{f}{Blended with CH$_3$OCH$_3$
($28_{5,23}$--$27_{6,22}$), $\nu_{\circ}$=230.368178 GHz;}
\tablenotetext{g}{Not listed in the JPL catalog, but included in
Cologne Database for Molecular Spectroscopy;}
\tablenotetext{h}{Blended with CH$_3$CHO $(12_{2,11}$--$11_{2,10})$,
$\nu_{\circ}$=230.315740 GHz;} \tablenotetext{i}{Blended with
CH$_3$OCHO $(31_{5,27}$--$31_{4,28})$, $\nu_{\circ}$=229.492067
GHz;} \tablenotetext{j}{Blended with CH$_3$OCH$_3$
$(10_{8,3}$--$11_{7,4})$, $\nu_{\circ}$=230.473591 GHz;}
\tablenotetext{k}{Blended with C$_2$H$_5$OH $(6_{5,2}$--$5_{4,2})$,
$\nu_{\circ}$=230.793864 GHz;} \tablenotetext{l}{Probably blended
with C$_3$H$_7$CN $(36_{6,31}$--$35_{5,30})$,
$\nu_{\circ}$=230.142063 GHz;} \tablenotetext{m}{Not identified.}
\end{deluxetable}

\end{document}